\documentclass[article]{article}

\usepackage[hmargin={15mm,15mm},vmargin={10mm,15mm}]{geometry}
\usepackage{amsmath,amsfonts,amssymb}
\usepackage{mathtools}
\usepackage{bm}
\usepackage{stmaryrd}
\usepackage{paralist}
\usepackage{booktabs}
\usepackage{calc}
\usepackage{hyperref}
\usepackage{pgfplots}
\usepackage{pgfplotstable}
\pgfplotsset{compat=1.16}
\usepackage{tikz}

\usepackage[normalem]{ulem}
\definecolor{col_p1}{RGB}{170,170,170}
\definecolor{col_p2}{RGB}{255,182,193}
\definecolor{col_p3}{RGB}{255,0,170}
\definecolor{col_p4}{RGB}{0,0,255}
\definecolor{col_p5}{RGB}{0,170,255}
\definecolor{col_p6}{RGB}{0,255,0}
\definecolor{col_p7}{RGB}{255,170,0}
\definecolor{col_p8}{RGB}{255,0,0}
\definecolor{col_p9}{RGB}{170,0,255}

\newcommand{\Real}{\mathbb{R}}

\newcommand{\tF}{t_{\rm F}}

\newcommand{\elemSymbol}[0]{T}
\newcommand{\faceSymbol}[0]{F}
\newcommand{\edgeSymbol}[0]{E}
\newcommand{\elem}[0]{T} 
\newcommand{\face}[0]{F} 
\newcommand{\edge}[0]{{\rm \edgeSymbol}}
\newcommand{\GhT}[0]{{\mathcal{G}}_\elem }
\newcommand{\GhV}[0]{{\bm{\mathfrak{g}}}_\elem }
\newcommand{\R}[0]{\mathbb{R}}
\newcommand{\Poly}[1]{\mathcal{P}^{#1}}
\newcommand{\Polyd}[2]{\mathbb{P}_{#1}^{#2}}

\newcommand{\sd}{sd}
\newcommand{\std}{std}
\newcommand{\td}{td}

\DeclareMathOperator{\CARD}{card}
\newcommand{\card}[1]{\CARD(#1)}
\newcommand{\Mh}{\mathcal{M}_h}
\newcommand{\Th}{\mathcal{\elemSymbol}_h}
\newcommand{\Fh}{\mathcal{\faceSymbol}_h}
\newcommand{\FT}{\mathcal{\faceSymbol}_\elem}
\newcommand{\normal}{\boldsymbol{n}}
\newcommand{\normalTF}{\boldsymbol{n}_{\elem \face}}
\newcommand{\npT}{\boldsymbol{n}_{\partial \elem}}

\newcommand{\dir}{{\rm D}}
\newcommand{\neu}{{\rm N}}
\newcommand{\internal}{{\rm i}}

\newcommand{\pTN}{\partial \elem^\neu}
\newcommand{\pTD}{\partial \elem^\dir}
\newcommand{\pTbullet}{\partial \elem^\bullet}

\newcommand{\sumFT}[0]{\sum_{\face \in \FT}}
\newcommand{\sumTh}[0]{\sum_{\elem \in \Th}}
\newcommand{\sumFTD}[0]{\sum_{\face \in \FT^{\dir}}}
\newcommand{\sumFTN}[0]{\sum_{\face \in \FT^\neu}}

\newcommand{\fb}{\bm{f}}
\newcommand{\pT}{{p}_\elem}
\newcommand{\pF}{{p}_\face}
\newcommand{\pTF}{\underline{p}_\elem}
\newcommand{\ppT}{p_{\partial \elem}}
\newcommand{\qT}{{q}_\elem}
\newcommand{\qpT}{{q}_{\partial \elem}}
\newcommand{\qF}{{q}_\face}
\newcommand{\qTF}{\underline{q}_\elem}
\newcommand{\uF}{\boldsymbol{u}_\face}
\newcommand{\uT}{\boldsymbol{u}_\elem}
\newcommand{\ueT}{\widehat{\boldsymbol{u}}_\elem}
\newcommand{\ueh}{\widehat{\boldsymbol{u}}_h}
\newcommand{\peh}{\widehat{p}_h}
\newcommand{\upT}{\boldsymbol{u}_{\partial \elem}}
\newcommand{\uVec}{\boldsymbol{u}}
\newcommand{\vVec}{\boldsymbol{v}}
\newcommand{\uTF}{\underline{\boldsymbol{u}}_\elem}
\newcommand{\vT}{\boldsymbol{v}_\elem}
\newcommand{\vpT}{\boldsymbol{v}_{\partial \elem}}
\newcommand{\vTF}{\underline{\boldsymbol{v}}_\elem}
\newcommand{\vF}{\boldsymbol{v}_\face}
\newcommand{\vb}[0]{\bm{v}}

\newcommand{\intT}[0]{\int_{\elem}}
\newcommand{\intpT}[0]{\int_{\partial\elem}}
\newcommand{\intpTN}{\int_{\partial\elem^\neu}}
\newcommand{\intpTD}{\int_{\partial\elem^\dir}}
\newcommand{\intF}[0]{\int_{\face}}

\newcommand{\lm}{{\rm slm}}
\newcommand{\ulm}{{\rm mnt}}
\newcommand{\cnt}{{\rm cnt}}
\newcommand{\pBC}{{\rm pBC}}

\newcommand{\tola}{tol_{a}}
\newcommand{\TimeEst} {r}
\newcommand{\lsnorm}[1]{\left\Vert #1 \right\Vert_{L^2(\Omega)}}

\newcommand{\eg}{e.g., }
\newcommand{\ea}{et al.}
\newcommand{\ie}{i.e., }

\newcommand{\Reynolds}{\mathrm{Re}}
\newcommand{\rbrackets}[1]{\left(#1\right)}
\newcommand{\ffunction}[2]{#1\rbrackets{#2}}

\newcommand{\MAT}[1]{\boldsymbol{#1}}
\newcommand{\VEC}[1]{{\sf #1}}
\newcommand{\UVEC}[1]{\underline{#1}}
\newcommand{\trans}{^\intercal}

\usepackage{amssymb}
\usepackage[affil-it]{authblk}
\newcommand{\email}[1]{\href{mailto:#1}{#1}}
\renewcommand\Affilfont{\footnotesize}

\begin{document}

  \title{Hybrid High-Order formulations with turbulence modelling capabilities for incompressible flow problems}
\author[1]{Lorenzo Botti}
\author[2]{Daniele A. Di Pietro}
\author[3]{Francesco Carlo Massa}
\affil[1]{Department of Engineering and Applied Sciences, University of Bergamo, Italy, \email{lorenzo.botti@unibg.it}}
\affil[2]{IMAG, Univ Montpellier, CNRS, Montpellier, France, \email{daniele.di-pietro@umontpellier.fr}}
\affil[3]{Department of Engineering and Applied Sciences, University of Bergamo, Italy, \email{francescocarlo.massa@unibg.it}}

\maketitle

  \begin{abstract}
    We propose a Hybrid High-Order (HHO) formulation of the incompressible Navier--Stokes equations,
    that is well suited to be employed for the simulation of turbulent flows.
    The spatial discretization relies on hybrid velocity and pressure spaces and the
    temporal discretization is based on Explicit Singly Diagonal Implicit Runge-Kutta (ESDIRK) methods.
    The formulation possesses some attractive features that can be fruitfully exploited
    when high-fidelity computations are required, namely:
    pressure-robustness, conservation of mass enforced cell-by-cell up to machine precision,
    robustness in the inviscid limit, implicit high-order accurate time stepping with local time step adaptation,
    reduced memory footprint thanks to static condensation of both velocity and pressure,
    possibility to exploit inherited $p$-multilevel solution strategies to improve performance of iterative solvers.
    After demonstrating the relevant properties of the scheme in practice, performing challenging 2D and 3D test cases,
    we consider the simulation of the Taylor--Green Vortex flow problem at Reynolds $1~600$.
  \end{abstract}

\section{Introduction}

In recent years, the introduction of Hybrid High-Order (HHO) and Hybridizable Discontinuous Galerkin (HDG) methods has provided a new ground to pursue the development of high-order accurate computational modelling tools for the simulation of incompressible turbulent flows.
In contrast to Discontinuous Galerkin methods, HHO and HDG methods are based on degrees of freedom that are broken polynomials on both the mesh and its skeleton.
Relevant features of hybrid schemes are:
\begin{inparaenum}[i)]
\item local (cell-by-cell) conservation of physical quantities,
\item increased convergence rates in the diffusion-dominated regime (see \cite{Cockburn.Di-Pietro.ea:16,Di-Pietro.Droniou.ea:15} for a discussion on this subject as well as a comparison between HHO and HDG),
\item robustness with mesh distortion and grading,
\item implicit time integration with reduced memory footprint of the Jacobian matrix.
\end{inparaenum}

This work contains a numerical investigation of the HHO scheme of \cite{BottiMassa22} applied to turbulent flows.
Unlike previous HHO schemes for incompressible flows (see, e.g., \cite{Aghili.Boyaval.ea:15,Di-Pietro.Ern.ea:16*1,Di-Pietro.Krell:18,BottiDiPietroDroniou2019}), the one considered relies on a hybrid discretization of both the velocity and pressure.
This choice, combined with an appropriate selection of local polynomial spaces, has been shown to be pressure-robust \cite{BBDiPietroMassa25} and, provided a suitable discretization of the time derivative and convective stabilization are adopted, Reynolds semi-robust \cite{Beirao-da-Veiga.Di-Pietro.ea:25}.
Following \cite{Linke:14}, pressure-robustness expresses the fact that the irrotational part of body forces only affects the pressure, leaving the velocity unaltered.
Pressure-robust HDG methods with hybrid pressure have been proposed in \cite{Rhebergen.Wells:18,Kirk2019}; see \cite[Remark~15]{BBDiPietroMassa25} and \cite[Remark~8]{Beirao-da-Veiga.Di-Pietro.ea:25} for a comparison with the method considered here or closely-related variations thereof.
Various strategies to obtain pressure-robustness and, possibly, Reynolds-semi-robustness in the framework of HHO methods have been studied in \cite{QUIROZ2020,QUIROZ2024,QUIROZ2025SM}, where the case of general polytopal meshes is also considered.
The aforementioned articles share in common the idea to reconstruct an $\boldsymbol{H}(\operatorname{div})$-conforming velocity approximation starting from non-conforming polynomial spaces.
Other research efforts to achieve pressure robustness based on $\boldsymbol{H}(\operatorname{div})$-conforming spaces for the velocity have been been made in \cite{LEHRENFELD2016,QiuZhao2024}.

In order to treat time-dependent problems in the convection-dominated flow regime, we combine here the upwind-stabilized variant of the space 
discretization of \cite{BottiMassa22} with high-order ESDIRK time integration schemes.
Efficient resolution of the algebraic problems relies on $p$-multigrid preconditioners in the spirit of \cite{Botti.Di-Pietro:22} 
and on a distributed memory implementation that is well suited to exploit modern HPC facilities. 
Numerical validation of the approach is carried out using classical two- and three-dimensional test cases.
Convergence rates are numerically evaluated for a wide range of Reynolds numbers, encompassing both the viscosity-dominated and convection-dominated regimes.
The behaviour of the discretization error is studied for the two-dimensional travelling waves and the three-dimensional Ethier--Steinman \cite{EthierSteinman94} solutions.
Robustness in the inviscid limit and numerical dissipation are assessed using the double shear layer problem.
Pressure-robustness is checked using the solution proposed in \cite{Lederer.Linke.ea:17}.
Finally, the capabilities of the proposed ESDIRK-HHO solution strategy are investigated targeting under-resolved turbulence modelling considering the Taylor--Green vortex problem \cite{TGVortexTaylor37} at Reynolds $1\,600$.

The paper is organized as follows.
In Sections~\ref{sec:INS} and~\ref{sec:HHOESDIRK} we respectively present the model problem and its discrete formulation.
A comprehensive numerical validation of the method is performed in Section~\ref{sec:numerical.validation} and an application to turbulence modelling is considered in Section~\ref{sec:turbulence}.
Finally, some conclusions are drawn in Section~\ref{sec:conclusion}.


\section{Continuous setting} \label{sec:INS}

Given a polygonal or polyhedral domain $\Omega\subset\Real^d$, $d\in\{2,3\}$, with boundary $\partial \Omega$, the initial divergence-free velocity field $\uVec_0: \Omega \rightarrow \Real^d$, and a finite time $\tF$,
the incompressible Navier--Stokes problem consists in finding the velocity field $\uVec: \Omega \times \lbrack 0,\tF\rbrack \rightarrow \Real^d$, and the pressure field $p: \Omega  \times (0,\tF\rbrack \rightarrow \Real$,
such that $\uVec(\cdot,0) = \uVec_0$ and
\begin{subequations}
  \label{nstokesProb}
  \begin{alignat}{2}\label{stokesProb:momentum}
    \frac{\partial \uVec}{\partial t} + \nabla \cdot \left[{(\uVec \otimes \uVec) -\nu \nabla \uVec }\right] + \nabla p &= \boldsymbol{f}
    &\qquad& \text{in $\Omega \times (0,\tF\rbrack$}, \\ \label{stokesProb:mass}
    \nabla \cdot \uVec &= 0
    &\qquad& \text{in $\Omega\times (0,\tF\rbrack$}, \\
    \uVec &= \boldsymbol{g}_\dir
    &\qquad& \text{on $\partial\Omega_\dir\times (0,\tF\rbrack$}. \\
    p \normal - \nu (\nabla \uVec)\normal  &= \boldsymbol{g}_\neu
    &\qquad& \text{on $\partial\Omega_\neu\times (0,\tF\rbrack$},
  \end{alignat}
\end{subequations}
where $\normal$ denotes the unit vector normal to $\partial\Omega$ pointing out of $\Omega$, $\nu > 0$ is the (constant) viscosity,
$\boldsymbol{g}_\dir$ and $\boldsymbol{g}_\neu$ denote, respectively, the prescribed velocity on the Dirichlet boundary $\partial\Omega_\dir\subset\partial\Omega$ and the prescribed traction on the Neumann boundary $\partial\Omega_\neu\coloneqq\partial\Omega\setminus\partial\Omega_\dir$,
while $\boldsymbol{f}:\Omega\to\Real^d$ is a given body force.


\section{ESDIRK-HHO discretization} \label{sec:HHOESDIRK}

\subsection{Mesh}
We consider meshes of the domain $\Omega$ corresponding to couples $\Mh\coloneqq(\Th,\Fh)$, where $\Th$
is a finite collection of polygonal (if $d = 2$) or polyhedral (if $d = 3$) cells
such that $h\coloneqq\max_{\elem\in\Th}h_\elem>0$ with $h_\elem$ denoting the diameter of $\elem$,
while $\Fh$ is a finite collection of polygons (if $d = 2$) or line segments (if $d = 3$) with $h_\face$ denoting the diameter of $\face$.
It is assumed henceforth that the mesh $\Mh$ is shape- and contact-regular, as detailed in \cite[Definition 1.4]{Di-Pietro.Droniou:20} and that its trace on $\partial \Omega$ is compatible with the partition $\partial \Omega = \partial \Omega_\dir \sqcup \partial \Omega_\neu$.
We respectively denote by $\Fh^\dir$ and $\Fh^\neu$ the sets of Dirichlet and Neumann faces.
For each mesh cell $\elem \in \Th$, the faces contained in the cell boundary $\partial T$ are collected in the set $\FT$ and we additionally let $\pTbullet \coloneqq \partial T \cap \partial \Omega_\bullet$ for $\bullet \in \{ \dir, \neu \}$.
For all $T\in\Th$ and all $F\in\FT$, $\normal_{TF}$ denotes the unit vector normal to $F$ pointing out of $T$ and $\npT:\partial T \to \Real^d$ is such that $\npT|_F \coloneqq \normal_{TF}$ for all $F\in\FT$.

\subsection{Discrete spaces}

Hybrid High-Order methods hinge on local polynomial spaces on mesh cells and faces.
For given integers $\ell\ge 0$ and $n\ge 1$, we denote by $\Polyd{n}{\ell}$ the space of $n$-variate polynomials of total degree $\le\ell$ (in short, of degree $\ell$).
For $X$ mesh cell or face, we denote by $\Poly{\ell}(X)$ the space spanned by the restriction to $X$ of functions in $\mathbb{P}_d^\ell$.

The global discrete spaces for the velocity and pressure unknowns are respectively defined as follows:
\[
\underline{\boldsymbol{V}}_h^{k}\coloneqq \left\{ \underline{\vVec}_h = \big( (\vVec_T)_{T\in\Th} , (\vVec_F)_{F \in \Fh} \big) :
\begin{array}{l}
  \text{${\vVec}_T \in \Poly{k+1}(T)^d$ for all $T\in\Th$},\\ \text{$\vVec_F  \in \Poly{k}(F)^d$ for all $F \in \Fh \setminus \Fh^\neu$,} \\ \text{$\vVec_F  \in \Poly{k+1}(F)^d$ for all $F \in \Fh^{\neu}$
  }
\end{array}
\right\}.
\]
\[
\underline{Q}_h^{k+1}\coloneqq\left\{
\underline{q}_h=\big((q_T)_{T\in\Th}, (q_F)_{F\in\Fh}\big)\,:\,
\text{$q_T\in\Poly{k}(T)$ for all $T\in\Th$ and $q_F\in\Poly{k+1}(F)$ for all $F\in\Fh$}
\right\}.
\]
The restriction of these spaces and their elements to a mesh cell $T \in \Th$ are denoted replacing the subscript $h$ by $T$ and are obtained collecting the components attached to $T$ and its boundary.
Given $\vTF \in \underline{\boldsymbol{V}}_T^{k}$, we will also denote by $\vVec_{\partial T}$ the function such that $\vVec_{\partial T}|_F = \vVec_F$ for all $F \in \FT$.
Given $\underline{q}_T \in  \underline{Q}_T^{k+1}$, the notation $q_{\partial T}$ is defined similarly.

\subsection{Gradient reconstruction operators}
For each mesh cell $\elem \in \Th$,
the tensor and vector gradient reconstruction operators
$\GhT^k : \underline{\boldsymbol{V}}_T^{k} \to \Poly{k}(\elem)^{d \times d}$
and $\GhV^{k+1} : \underline{Q}_T^{k+1} \to \Poly{k+1}(\elem)^{d}$
are respectively defined as follows:
For all $(\vTF, \qTF) \in \underline{\boldsymbol{V}}_T^{k} \times \underline{Q}_T^{k+1}$,
\begin{gather*} 
  \intT \GhT^k \vTF : \bm{\tau}
  = \intT \nabla \vT : \bm{\tau}
  - \intpT (\vT - \vpT) \otimes \npT :  \bm{\tau}
  \qquad \forall \bm{\tau} \in \Poly{k}(\elem)^{d \times d},
  \\ 
  \intT \GhV^{k+1} {\qTF} \cdot \bm{\vb}   
  = - \intT \qT \;  \nabla \cdot \bm{\vb}
  + \intpT \qpT \bm{\vb} \cdot \npT
  \qquad \forall \bm{\vb} \in \Poly{k+1}(\elem)^{d}.
\end{gather*}

\subsection {Semi-discrete HHO formulation} \label{sec:SpaceDiscr}
We first present the semi-discrete formulation of the incompressible Navier--Stokes problem in \eqref{nstokesProb} omitting
the terms that are responsible of the imposition of boundary conditions. We remark that the resulting formulation
is well suited to tackle the imposition of periodic boundary conditions on $\partial \Omega$,
since, in this case, $\Fh^\dir = \Fh^\neu = \varnothing$, all faces are internal.
The application to periodic computational domains will be often encountered in the test cases proposed in the numerical results section.
Notice that a proper handling of internal faces laying on periodic boundaries must be ensured.
For example, once periodic boundaries couples are defined, assuming that their faces are properly matched,
periodic faces couples can be identified and each pair of faces can treated as a unique internal face.

Given $(\uTF, \pTF)\in \underline{\boldsymbol{V}}_T^{k}\times \underline{Q}_T^{k+1}$, the local spatial discretizations
$\sd^\lm_{T,\pBC} ((\uTF,\pTF);\cdot):\underline{\boldsymbol{V}}_T^{k}\to\Real$ of the steady momentum equation and $ \sd^\cnt_{T,\pBC}(\uTF;\cdot):\underline{Q}^{k+1}_T\to\Real$ of the continuity equations for periodic boundary conditions are such that,
for all $\vTF\in\underline{\boldsymbol{V}}_T^{k}$ and all $\qTF\in\underline{Q}_T^{k+1}$,
\begin{alignat}{1}
  \sd^{\lm}_{T,\pBC} ((\uTF,\pTF); \vTF) \coloneqq &
    - \int_T (\uT \otimes \uT) : \nabla {\bm{\pi}_{T}^k}\vT
    + \intpT \left[
      \bigl(\uT \cdot \npT \bigr)^{\oplus} \uT
      + \bigl(\uT \cdot \npT \bigr)^{\ominus} \upT
      \right] \cdot \bigl({\bm{\pi}_{T}^k} \vT - {\bm{\pi}_{\partial T}^k} \vpT\bigr) \label{lmConvHHO}\\
    & + \intT \nu \, \GhT^k \uTF : \GhT^k \vTF
    + \intpT \frac{\nu}{h_T} \, \bm{\pi}_{\partial T}^k(\uT - \upT) \cdot \bm{\pi}_{\partial T}^k(\vT - \vpT)  \label{lmDiffHHO}\\    
    &+ \intT \GhV^{k+1} \pTF \cdot \vT  - \intT \fb \cdot \vT \nonumber, 
  \\
  \sd^{\cnt}_{T,\pBC} (\uTF;\qTF) \coloneqq &\intT \GhV^{k+1} \qTF \cdot \uT \nonumber,
\end{alignat}
where $\alpha^{\oplus} \coloneqq \frac{1}{2} (\alpha + |\alpha|)$ and $\alpha^{\ominus} \coloneqq \frac{1}{2} (\alpha - |\alpha|)$,
$\bm{\pi}_{\partial T}^k$ is the $L^2$-orthogonal projector onto the broken polynomial space $\left\{ \vVec \in L^2(\partial T)^d \;:\; \text{$\vVec|_F \in \Poly{k}(F)^d$ for all $F \in \FT$} \right\}$ and 
{$\bm{\pi}_{T}^k$ is the $L^2$-orthogonal projector onto $\Poly{k}(T)^d$}.
{ The idea of introducing $L^2$-projections of test functions in the convective term formulation of \eqref{lmConvHHO} is inspired by \cite{Beirao-da-Veiga.Di-Pietro.ea:25}
and improves the performance of the method in the convection-dominated regime.
As will be demonstrated in the numerical results section, the resulting HHO formulation is Reynolds semi-robust, 
meaning that velocity convergence rates observed in the convection-dominated regime 
are half an order suboptimal with respect to the convergence rates observed in the diffusion-dominated regime. 
For the sake of comparison, omitting $L^2$-projections in the convective term formulation 
would endow losing a full order of convergence in the convection-dominated regime, see \cite{Beirao-da-Veiga.Di-Pietro.ea:25} for details. 
The diffusion term stabilization of \eqref{lmDiffHHO}, also involving $L^2$-projection operators, 
was first proposed by \cite{Lehrenfeld:10}, see also \cite{Cockburn.Di-Pietro.ea:16}.}

For Dirichlet and Neumann boundary conditions, on the other hand, given $(\uTF, \pTF)\in \underline{\boldsymbol{V}}_T^{k}\times \underline{Q}_T^{k+1}$, the local spatial discretizations
$\sd^\lm_{T} ((\uTF,\pTF);\cdot):\underline{\boldsymbol{V}}_T^{k}\to\Real$ of the steady linear momentum
and $ \sd^\cnt_{T}(\uTF;\cdot):\underline{Q}^{k+1}_T\to\Real$ of the continuity equations are respectively such that,
for all $\vTF\in\underline{\boldsymbol{V}}_T^{k}$ and all $\qTF\in\underline{Q}_T^{k+1}$,
\begin{alignat}{1}
  \sd^{\lm}_{T} \bigl((\uTF,\pTF); \vTF\bigr) \coloneqq& \sd^{\lm}_{T,\pBC} ((\uTF,\pTF); \vTF) \nonumber \\
  &
  + \intpTN \bigl( \upT \cdot \npT \bigr)  \; \bigl(\upT \cdot {\bm{\pi}_{\partial T}^k}\vpT \bigr)
  - \intpTN \ppT (\vpT \cdot \npT)
  + \intpTN \bm{g}_\neu \cdot \vpT \label{lmNeu} \\
  &
  + \intpTD \bigl( \upT \cdot \npT \bigr)^{\oplus} \; (\upT \cdot \vpT)
  + \intpTD \bigl( \bm{g}_\dir \cdot \npT \bigr)^{\ominus} \; (\bm{g}_\dir \cdot \vpT) \label{lmDirConv}\\
  &
  + \intpTD  \bigl( (\upT - \bm{g}_\dir) \otimes \npT \bigr) : \left( \nu \GhT \vTF + \frac{\nu}{h_\face} \vpT \otimes \npT \right)
  - \intpTD \nu \GhT \uTF : (\vpT \otimes \npT ), \label{lmDirDiff}
  \\
  \sd^{\cnt}_{T} (\uTF;\qTF)
  \coloneqq & \sd^{\cnt}_{T,\pBC} (\uTF;\qTF)
  - \intpTD (\bm{g}_\dir \cdot \npT) \, \qpT
  - \intpTN (\upT \cdot \npT) \, \qpT. \nonumber
\end{alignat}
The terms in \eqref{lmDirConv} and \eqref{lmDirDiff} are responsible for the weak imposition of Dirichlet boundary conditions and pertain to
the spatial discretization of the convective term and the viscous term, respectively, see \cite[Remark 6]{BottiDiPietroDroniou2019}.
The terms in \eqref{lmNeu} are responsible for the imposition of Neumann boundary conditions: the first and the second term, respectively, pertain to the spatial discretization of the convective term and the pressure gradient.

Given $(\uTF, \pTF)\in \underline{\boldsymbol{V}}_T^{k}\times \underline{Q}_T^{k+1}$,
the local spatial discretization of the unsteady momentum equation is such that,
for all $\vTF\in\underline{\boldsymbol{V}}_T^{k}$,
$$
\sd_{T}^\ulm\bigl((\uTF,\pTF); \vTF\bigr) \coloneqq
\intT \frac{\partial \uT}{\partial t} \cdot \vT + {\intpT \left( \frac{\partial \upT}{\partial t} - \frac{\partial \uT}{\partial t} \right) \cdot (\vpT - \vT)} + \sd^\lm_{T}\left((\underline{\boldsymbol{u}}_T,\underline{p}_T);\underline{\boldsymbol{v}}_T\right).
$$
{ We remark that the second term in the expression above is a time derivative stabilization designed to enhance the robustness of the formulation in the case of vanishing viscosity.}

The global residuals of the spatial discretization
$r^\ulm_{h}((\underline{\boldsymbol{u}}_h,\underline{p}_h);\cdot):\underline{\boldsymbol{V}}^{k}_h\to\Real$ and
$r^\cnt_{h}(\underline{\boldsymbol{u}}_h;\cdot):\underline{Q}^{k+1}_h\to\Real$ are obtained by cell-by-cell assembly of the local discretizations, \textit{i.e.}:
For all $\underline{\boldsymbol{v}}_h\in\underline{\boldsymbol{V}}_h^{k}$ and all $\underline{q}_h\in\underline{Q}_h^{k+1}$,
\begin{equation}
  \label{eq:HHOdiscrGlobalResHDIV}
  r_{h}^\ulm\left((\underline{\boldsymbol{u}}_h,\underline{p}_h);\underline{\boldsymbol{v}}_h\right)
  \coloneqq \sum_{T\in\Th}  \sd^\ulm_{T}\left((\underline{\boldsymbol{u}}_T,\underline{p}_T);\underline{\boldsymbol{v}}_T\right),\qquad
  r_{h}^\cnt(\underline{\boldsymbol{u}}_h;\underline{q}_h) \coloneqq \sum_{T\in\Th} \sd^\cnt_{T}(\underline{\boldsymbol{u}}_T;\underline{q}_{T}).
\end{equation}
The global problem reads: Find $(\underline{\boldsymbol{u}}_h,\underline{p}_h)\in\underline{\boldsymbol{V}}_h^{k}\times\underline{Q}^{k+1}_h$ such that
\begin{equation}\label{eq:hhoGlobalProblem}
  \begin{alignedat}{2}
    r_{h}^\ulm((\underline{\boldsymbol{u}}_h,\underline{p}_h);\underline{\boldsymbol{v}}_h) &= 0 &\qquad& \forall \underline{\boldsymbol{v}}_h\in\underline{\boldsymbol{V}}_h^{k},
    \\
    r_{h}^\cnt(\underline{\boldsymbol{u}}_h;\underline{q}_h) &= 0 &\qquad& \forall \underline{q}_h\in\underline{Q}_h^{k+1}.
  \end{alignedat}
\end{equation}

\subsection{ESDIRK temporal discretization} \label{sec:TempDiscr}

We describe hereafter the time stepping algorithm focusing, for the sake of simplicity, on the case of periodic boundary conditions.
Based on multistage ESDIRK time marching strategies, the solutions $\boldsymbol{u}_h |_{t+\delta t}, {p}_h |_{t+\delta t}$
are computed considering an approximated time integration formula for the global residuals in \eqref{eq:HHOdiscrGlobalResHDIV} over the time interval $\delta t$:
\begin{equation}
\label{eq:HHOdiscrTemp}
\begin{aligned}
\sum_{T\in\Th}  \intT \frac{\delta \uT}{\delta t} \cdot \vT 
+ {\sum_{T\in\Th}  \intpT \left(\frac{\delta \upT}{\delta t} - \frac{\delta \uT }{\delta t}\right) \cdot (\vpT - \vT) }
+  \sum_{T\in\Th} \sum_{i = 1}^{s} b_i \, \sd^\lm_{T}\left((\uTF|_{t+\delta t_i},\pTF|_{t+\delta t_i});\underline{\boldsymbol{v}}_T\right) &= 0, \\
\sum_{T\in\Th} \sum_{i = 1}^{s} b_i \, \sd^\cnt_{T}(\uTF|_{t+\delta t_i};\qTF) &= 0,
\end{aligned}
\end{equation}
where {$\delta \uVec_X \coloneqq \uVec_X |_{t+\delta t} - \uVec_X |_t$}, and, for all $T \in \Th$,  $\uTF|_{t+\delta t_i}$ and $\pTF|_{t+\delta t_i}$ are the solutions of the ESDIRK $i$th-stage,
with $\delta t_i = \delta t \; \sum_{j=1}^i a_{ij}$.
The number of stages $s$ and the real coefficients $a_{ij}$, with $i = 1,\dots,s$ and $j = 1,\dots,i$, uniquely define an ESDIRK formulation.
In particular, for all schemes belonging to the ESDIRK family, it holds that $a_{11}=0$ and $b_i = a_{si}$, for $i = 1,\dots,s$.
As a consequence, since $\delta t_1 = 0 $ and $\delta t_s = \delta t$,
the approximated time integration formula in \eqref{eq:HHOdiscrTemp} is equivalent to the last ESDIRK stage
and $\boldsymbol{u}_h|_{t+\delta t}, {p}_h|_{t+\delta t}$ is the last stage solution.
The stages solution are built incrementally, starting from the first stage, moving to second, and so on up to the last stage.
In this work we rely on third, fourth and fifth order accurate ESDIRK schemes requiring 
four, six and eight stages, respectively, see~\cite{KENNEDY2003139,Kennedy2016} for details.
To complete the presentation of the time marching strategy we need to specify how the stages solutions are computed.

Let's introduce the ESDIRK $i$th-stage formulation.
Given $(\uTF|_{t+\delta t_j}, \pTF|_{t+\delta t_j}) \in \underline{\boldsymbol{V}}_T^{k}\times \underline{Q}_T^{k+1}$, with $j=1,\dots,i$,
the local spatial and temporal discretizations
$\std^\lm_{i,T} ((\uTF(t),\pTF(t));\cdot):\underline{\boldsymbol{V}}_T^{k}\to\Real$ of the momentum equation
and $ \std^\cnt_{i,T}(\uTF(t);\cdot):\underline{Q}^{k+1}_T\to\Real$ of the continuity equations at the $i$th-stage are such that,
for all $\vTF\in\underline{\boldsymbol{V}}_T^{k}$ and all $\qTF\in\underline{Q}_T^{k+1}$,
\begin{subequations}
\label{eq:HHOdiscrSpaceTemp}
\begin{alignat}{1}
 \std^\ulm_{i,T}\left((\uTF(t),\pTF(t));\vTF \right) &\coloneqq 
 \intT \frac{\delta_i \uT}{\delta t} \cdot \vT 
+ { \intpT \left(\frac{\delta_i \upT}{\delta t} - \frac{\delta_i \uT }{\delta t}\right) \cdot (\vpT - \vT) } +
                                 \sum_{j = 1}^{i} a_{ij} \, \sd^\lm_{T}\left((\uTF|_{t+\delta t_j},\pTF|_{t+\delta t_j});\underline{\boldsymbol{v}}_T\right)  \\
 \std^\cnt_{i,T} (\uTF(t);\qTF)&\coloneqq \sum_{j = 1}^{i} a_{ij} \, \sd^\cnt_{T}(\uTF|_{t+\delta t_j};\qTF)
\end{alignat}
\end{subequations}
where {$\delta_i \uVec_X \coloneqq \uVec_X |_{t+\delta t_i} - \uVec_X |_t$}.
We remark that, the spatial and temporal discretizations $\std^\ulm(\cdot,\cdot), \std^\cnt(\cdot,\cdot)$ 
defined in \eqref{eq:HHOdiscrTemp}-\eqref{eq:HHOdiscrSpaceTemp} are formulated omitting the dependence from the boundary conditions and forcing terms. 
In doing so, we implicitly assume that $\boldsymbol{f}$ (and also $\boldsymbol{g}_\neu, \boldsymbol{g}_\dir $ when non-periodic boundary conditions are considered) is evaluated at the same time points as the velocity and pressure variables $\uTF, \pTF$.

The global residuals of the ESDIRK $i$th-stage
 $r^\ulm_{i,h}\left((\underline{\boldsymbol{u}}_h(t),\underline{p}_h(t));\cdot\right):\underline{\boldsymbol{V}}^{k}_h\to\Real$, and
 $r^\cnt_{i,h}\left(\underline{\boldsymbol{u}}_h (t);\cdot\right):\underline{Q}^{k+1}_h\to\Real$
are obtained by cell-by-cell assembly of the local spatial and temporal discretizations, \textit{i.e.}:
For all $\underline{\boldsymbol{v}}_h\in\underline{\boldsymbol{V}}_h^{k}$ and all $\underline{q}_h\in\underline{Q}_h^{k+1}$,
\begin{subequations}
\begin{alignat}{1}
r_{i,h}^\ulm\left((\underline{\boldsymbol{u}}_h(t),\underline{p}_h(t));\underline{\boldsymbol{v}}_h\right)
 &\coloneqq \sum_{T\in\Th}  \std^\ulm_{i,T}\left((\underline{\boldsymbol{u}}_T(t),\underline{p}_T(t));\underline{\boldsymbol{v}}_T\right), \nonumber \\
r_{i,h}^\cnt(\underline{\boldsymbol{u}}_h(t);\underline{q}_h) &\coloneqq \sum_{T\in\Th} \std^\cnt_{i,T}(\underline{\boldsymbol{u}}_T(t);\underline{q}_{T}). \nonumber
\end{alignat}
\end{subequations}

The ESDIRK $i$th-stage problem, that is the time discrete counterpart of problem \eqref{eq:hhoGlobalProblem},
reads:
Given $\boldsymbol{u}_h|_{t+\delta t_j},p_h|_{t+\delta t_j}$, with $j=1,...,i-1$, find $\boldsymbol{u}_h|_{t+\delta t_i}, {p}_h|_{t+\delta t_i}$ such that
\begin{equation}
  \label{HHOdiscrTempResidualStagei}
  \text{%
    ${r}_{i,h}^\ulm\left((\underline{\boldsymbol{u}}_h(t),\underline{p}_h(t));\underline{\boldsymbol{v}}_h\right) = 0 \; \; \forall \underline{\boldsymbol{v}}_h\in\underline{\boldsymbol{V}}_h^{k},$
    \quad and \quad
    ${r}_{i,h}^\cnt\left(\underline{\boldsymbol{u}}_h(t);\underline{q}_h\right) = 0 \; \; \forall \underline{q}_h\in\underline{Q}_h^{k+1}$.
  }
\end{equation}
The solution of the above problem can be sought by means of Newton's method.
Introducing the shortcut notations
$\vVec_T \gets (\vVec_T, (\boldsymbol{0})_{F\in\FT})$
and $\vVec_{\partial T} \gets (\boldsymbol{0}, (\vVec_F)_{F\in\FT})$ in the second argument of $\std^\ulm_{i,T}$,
and $q_T \gets (q_T,(0)_{F\in\FT})$ and
$q_{\partial T} \gets (0,(q_F)_{F\in\FT})$ in the second argument of $\std^\cnt_{i,T}$,
the Newton's algorithm for solving \eqref{HHOdiscrTempResidualStagei} reads:
Find $\delta \underline{\uVec}_h, \delta \underline{p}_h$ such that
\begin{equation}
\label{eq:newtonCont}
\begin{aligned}
 & -\sum_{\elem \in \Th}
  \left(\begin{bmatrix}
   \std^\ulm_{i,T}\left((\underline{\boldsymbol{u}}_T(t),\underline{p}_T(t));\boldsymbol{v}_T\right)  \\
   \std^\ulm_{i,T}\left((\underline{\boldsymbol{u}}_T(t),\underline{p}_T(t));\boldsymbol{v}_{\partial T}\right) \\
   \std^\cnt_{i,T}\left(\underline{\boldsymbol{u}}_T(t);{q}_T\right)  \\
   \std^\cnt_{i,T}\left(\underline{\boldsymbol{u}}_T(t);{q}_{\partial T}\right)
  \end{bmatrix} \right) \\
&\qquad \qquad =\sum_{\elem \in \Th}
  \left(\begin{bmatrix}
    \frac{\partial \std^\ulm_{i,T}\left((\uTF(t),\pTF(t));\vT \right)}{\partial \uT|_{t+\delta t_i}}
  & \frac{\partial \std^\ulm_{i,T}\left((\uTF(t),\pTF(t));\vT \right)}{\partial \upT|_{t+\delta t_i}}
  & \frac{\partial \std^\ulm_{i,T}\left((\uTF(t),\pTF(t));\vT \right)}{\partial \pT|_{t+\delta t_i}}
  & \frac{\partial \std^\ulm_{i,T}\left((\uTF(t),\pTF(t));\vT \right)}{\partial \ppT|_{t+\delta t_i}}    \\
    \frac{\partial \std^\ulm_{i,T}\left((\uTF(t),\pTF(t));\vpT \right)}{\partial \uT|_{t+\delta t_i}}
  & \frac{\partial \std^\ulm_{i,T}\left((\uTF(t),\pTF(t));\vpT \right)}{\partial \upT|_{t+\delta t_i}}
  & \frac{\partial \std^\ulm_{i,T}\left((\uTF(t),\pTF(t));\vpT \right)}{\partial \pT|_{t+\delta t_i}}
  & \frac{\partial \std^\ulm_{i,T}\left((\uTF(t),\pTF(t));\vpT \right)}{\partial \ppT|_{t+\delta t_i}}    \\
    \frac{\partial \std^\cnt_{i,T}\left(\uTF(t);\qT \right)}{\partial \uT|_{t+\delta t_i}}
  & \frac{\partial \std^\cnt_{i,T}\left(\uTF(t);\qT \right)}{\partial \upT|_{t+\delta t_i}}
  & \frac{\partial \std^\cnt_{i,T}\left(\uTF(t);\qT \right)}{\partial \pT|_{t+\delta t_i}}
  & \frac{\partial \std^\cnt_{i,T}\left(\uTF(t);\qT \right)}{\partial \ppT|_{t+\delta t_i}}    \\
    \frac{\partial \std^\cnt_{i,T}\left(\uTF(t);\qpT \right)}{\partial \uT|_{t+\delta t_i}}
  & \frac{\partial \std^\cnt_{i,T}\left(\uTF(t);\qpT \right)}{\partial \upT|_{t+\delta t_i}}
  & \frac{\partial \std^\cnt_{i,T}\left(\uTF(t);\qpT \right)}{\partial \pT|_{t+\delta t_i}}
  & \frac{\partial \std^\cnt_{i,T}\left(\uTF(t);\qpT \right)}{\partial \ppT|_{t+\delta t_i}}    \\
  \end{bmatrix}
  \begin{bmatrix}
    \delta \uT \\
    \delta \upT \\
    \delta \pT\\
    \delta \ppT
  \end{bmatrix} \right)
\end{aligned}
\end{equation}
  replace
  $\underline{\uVec}_h|_{t+\delta t_i} \gets \underline{\uVec}_h|_{t+\delta t_i} + \delta \underline{\uVec}_h, 
  \underline{p}_h|_{t+\delta t_i} \gets \underline{p}_h|_{t+\delta t_i} + \delta \underline{p}_h$
until $\delta \underline{\uVec}_h, \delta \underline{p}_h$ is small enough.
The local Jacobian matrix on the left hand side can be computed cell-by-cell and assembled into the global Jacobian matrix.
Nevertheless, in order to improve efficiency of the solution strategy and reduce the memory footprint of ESDIRK-HHO formulations, static condensation strategies can be fruitfully exploited, as described in the next section.

\subsection{Algebraic expression for Newton's method and static condensation} \label{sec:StaticCond}
The unknowns for a mesh cell $\elem\in\Th$ correspond to the coefficients of the expansions of the velocity and pressure in the selected local bases.
Assuming that the unknowns are ordered so that cell velocities come first and boundary velocities next, these coefficients are collected in the vectors
\begin{equation*}
  \text{
    $\UVEC{U}_T=\begin{bmatrix} \VEC{U}_T \\ \VEC{U}_{\partial T} \end{bmatrix}$
    and $\UVEC{P}_T=\begin{bmatrix} \VEC{P}_T \\ \VEC{P}_{\partial T} \end{bmatrix}$,
  }
\end{equation*}
where the block partition is the one naturally induced by the selected ordering of unknowns.

Let's introduce the vector representations
$\UVEC{D}_{T}^\ulm=\begin{bmatrix}\VEC{D}_{T}^\ulm\\ \VEC{D}_{\partial T}^\ulm\end{bmatrix}$
and $\UVEC{D}_{T}^\cnt=\begin{bmatrix}\VEC{D}_{T}^\cnt\\ \VEC{D}_{\partial T}^\cnt\end{bmatrix}$
of the local spatial and temporal discretizations
$\std^\ulm_{i,T}\left((\uTF(t),\pTF(t));\vTF \right), \, \std^\cnt_{i,T} (\uTF(t);\qTF)$ defined in \eqref{eq:HHOdiscrSpaceTemp}.
The block partition is the one induced by mimicking the selected ordering of unknowns for the expansion bases.

Let's also introduce the matrix representation of the relevant blocks of the local Jacobian matrix in \eqref{eq:newtonCont}
\begin{align*}
  \begin{bmatrix}
    \MAT{A}_{T  T} & \MAT{A}_{T  \partial T}  \\
    \MAT{A}_{\partial T T} & \MAT{A}_{\partial T  \partial T}
  \end{bmatrix}
  &\sim
  \begin{bmatrix}
    \frac{\partial \std^\ulm_{i,T}\left((\uTF(t),\pTF(t));\vT \right)}{\partial  \uT|_{t+\delta t_i}}
    & \frac{\partial \std^\ulm_{i,T}\left((\uTF(t),\pTF(t));\vT \right)}{\partial  \upT|_{t+\delta t_i}}   \\
    \frac{\partial \std^\ulm_{i,T}\left((\uTF(t),\pTF(t));\vpT \right)}{\partial \uT|_{t+\delta t_i}}
    & \frac{\partial \std^\ulm_{i,T}\left((\uTF(t),\pTF(t));\vpT \right)}{\partial \upT|_{t+\delta t_i}}
  \end{bmatrix},
  \\
  \begin{bmatrix}
    \MAT{B}_{T  T} & \MAT{0}  \\
    \MAT{B}_{\partial T T} & \MAT{B}_{\partial T  \partial T}
  \end{bmatrix}
  &\sim
  \begin{bmatrix}
    \frac{\partial \std^\cnt_{i,T}\left(\uTF(t);\qT \right)}{\partial \uT|_{t+\delta t_i}}
    & \frac{\partial \std^\cnt_{i,T}\left(\uTF(t);\qT \right)}{\partial \upT|_{t+\delta t_i}} \\
    \frac{\partial \std^\cnt_{i,T}\left(\uTF(t);\qpT \right)}{\partial \uT|_{t+\delta t_i}}
    & \frac{\partial \std^\cnt_{i,T}\left(\uTF(t);\qpT \right)}{\partial \upT|_{t+\delta t_i}}
  \end{bmatrix}.
\end{align*}
The matrices on the left are obtained from the terms on the right
by plugging local expansions and bases in the spatial-temporal discretizations.
Notice that also the derivatives with respect to velocity and pressure
should be replaced with derivatives computed with
respect to the coefficients of the expansions in the corresponding local bases.

According to the notation above, the algebraic counterpart of problem \eqref{eq:newtonCont} reads
\begin{equation*}
  -\sum_{\elem \in \Th}\left(\begin{bmatrix}
    \VEC{D}^\ulm_{T} \\
    \VEC{D}^\ulm_{\partial T} \\
    \VEC{D}^{\cnt}_{T} \\
    \VEC{D}^{\cnt}_{\partial T}
  \end{bmatrix}\right)
  =
  \sum_{\elem \in \Th}\left(\begin{bmatrix}
    \MAT{A}_{T T}           & \MAT{A}_{T\partial T}            & \MAT{B}_{T  T}\trans  & \MAT{B}_{\partial T T}\trans    \\
    \MAT{A}_{\partial T  T} & \MAT{A}_{\partial T  \partial T} & \MAT{0}               & \MAT{B}_{\partial T  \partial T}\trans \\
    \MAT{B}_{T  T}          & \MAT{0}                          & \MAT{0}               & \MAT{0} \\
    \MAT{B}_{\partial T T}  & \MAT{B}_{\partial T  \partial T} & \MAT{0}               & \MAT{0}
  \end{bmatrix}
  \begin{bmatrix}
    \VEC{\delta U}_T \\
    \VEC{\delta U}_{\partial T} \\
    \VEC{\delta P}_T \\
    \VEC{\delta P}_{\partial T}
  \end{bmatrix} \right)
\end{equation*}
and, rearranging the unknowns and equations, the local problem can be rewritten as follows
\begin{equation}\label{eq:zero-residual.TII}\renewcommand{\arraystretch}{1.3}
  -\sum_{\elem \in \Th}\left(\left[\begin{array}{c}
    \VEC{D}^\ulm_T \\
    \VEC{D}^\ulm_T \\ 
    \VEC{D}^\cnt_{\partial T} \\
    \VEC{D}^\cnt_{\partial T}
  \end{array}\right] \right)
  =
  \sum_{\elem \in \Th}\left(\left[\begin{array}{cc|cc}
    \MAT{A}_{T T}           & \MAT{B}_{T  T}\trans             & \MAT{A}_{T\partial T}                        & \MAT{B}_{\partial T T}\trans    \\
    \MAT{B}_{T T}           & \MAT{0}                          & \MAT{0}                                      & \MAT{0}   \\
    \MAT{A}_{\partial T  T} & \MAT{0}                          & \MAT{A}_{\partial T  \partial T}             & \MAT{B}_{\partial T  \partial T}\trans \\
    \MAT{B}_{\partial T  T} & \MAT{0}                          & \MAT{B}_{\partial T  \partial T}             & \MAT{0}
    \end{array}
  \right]
  \left[\begin{array}{c}
    \VEC{\delta U}_T \\
    \VEC{\delta P}_T \\ 
    \VEC{\delta U}_{\partial T} \\
    \VEC{\delta P}_{\partial T}
  \end{array}\right] \right).
\end{equation}
The only unknowns that are globally coupled are those collected in the subvector
$\begin{bmatrix}
  \VEC{\delta U}_{\partial T} \\
  \VEC{\delta P}_{\partial T}
\end{bmatrix}$,
while the remaining unknowns collected in
$\begin{bmatrix}
  \VEC{\delta U}_T \\
  \VEC{\delta P}_T
\end{bmatrix}$
can be eliminated by expressing them in terms of the former.
After performing this local elimination, the algebraic counterpart of the Newton method reads
\begin{equation*}
  - \sum_{\elem \in \Th}\left(\begin{bmatrix}
    \VEC{D}^\ulm_{\partial T} \\
    \VEC{D}^\cnt_{\partial T}
  \end{bmatrix}
  - \begin{bmatrix}
    \MAT{A}_{\partial T T} & \MAT{0} \\
    \MAT{B}_{\partial T T} & \MAT{0}
  \end{bmatrix}
  \begin{bmatrix}
    \MAT{A}_{T T} & \MAT{B}_{TT}\trans  \\
    \MAT{B}_{T  T} & \MAT{0}
  \end{bmatrix}^{-1}
  \begin{bmatrix}
    \VEC{D}^\ulm_{T} \\
    \VEC{D}^\cnt_{T}
  \end{bmatrix} \right)
 =
  \sum_{\elem \in \Th}\left(\MAT{S}_T^{}\begin{bmatrix}
  \VEC{\delta U}_{\partial T} \\
  \VEC{\delta P}_{\partial T}
  \end{bmatrix}\right),
\end{equation*}
where $\MAT{S}_T^{}$ denotes the Schur complement of the top left block of the left-hand side matrix in \eqref{eq:zero-residual.TII}, that is,
\begin{equation*}
  \renewcommand{\arraystretch}{1.3}
  \MAT{S}_T^{}
  =
    \MAT{S}_{\partial T \partial T}^{}
  \coloneqq \begin{bmatrix}
    \MAT{A}_{\partial T  \partial T} & \MAT{B}_{\partial T \partial T}\trans
    \\
    \MAT{B}_{\partial T \partial T} & \MAT{0}
  \end{bmatrix} -
   \begin{bmatrix}
    \MAT{A}_{\partial T  T} & \MAT{0}
    \\
    \MAT{B}_{\partial T T} & \MAT{0}
  \end{bmatrix}\begin{bmatrix}
    \MAT{A}_{T T} & {\MAT{B}}_{TT}\trans
      \\
      {\MAT{B}}_{T  T} & \MAT{0}
  \end{bmatrix}^{-1}\begin{bmatrix}
    \MAT{A}_{T\partial T} & \MAT{B}_{\partial T  T}\trans
    \\
    \MAT{0} & \MAT{0}
  \end{bmatrix}.
\end{equation*}
When increasing the polynomial degree $k$, the dimension of polynomial spaces over mesh cells grows much faster that the dimension of polynomial spaces over mesh faces.
Accordingly, the possibility to statically condense the cells unknowns
is the key for achieving efficiency when high-order accurate formulation are employed.
In order to further improve the performance of the iterative linear solver required for the solution
of each Newton step we rely on the $p$-multilevel preconditioner strategy proposed in \cite{BottiDiPietroHHOpMG2021}.


\subsection{Time step adaptation}
Local time step adaptation enables step-by-step control of the time integration error. 
The key element of the adaptation procedure is an estimator of the temporal discretization error, 
thus, based on the temporal error metric and a user defined threshold tolerance, the time step can be increased or reduced
and the solution can be accepted or recomputed in order to meet the prescribed precision.  

Let $\uVec_h|_{t} \in L^2(\Omega)^d$ and $p_h|_{t} \in L^2(\Omega)$ 
be the discrete velocity and pressure solution at time $t$, obtained by gluing together the cell components.
The time integration error is computed as:
\begin{equation*}
\TimeEst = \lsnorm{\uVec_h|_{t+\delta t} -\ueh|_{t+\delta t}} + \lsnorm{p_h|_{t+\delta t} -\peh|_{t+\delta t}},
\end{equation*}
where both $\uVec_h|_{t+\delta t}, p_h|_{t+\delta t}$, and $\ueh|_{t+\delta t},\peh|_{t+\delta t}$ are 
computed by means of a $s$-stages ESDIRK time integration strategy, but the less accurate solution $\ueh|_{t+\delta t},\peh|_{t+\delta t}$ 
is obtained by replacing the ESDIRK coefficients $b_{i}$ in \eqref{eq:HHOdiscrTemp} with a different set of coefficients $\widehat{b}_{i}$, for $i=1,\dots,s$.
In the ESDIRK nomenclature, $\ueh|_{t+\delta t},\peh|_{t+\delta t}$ is referred as the \emph{embedded} solution.

Following \cite{Soderlind:2003,Soderlind:2006}, 
at the end of each ESDIRK stage, we require that	
\begin{equation}\label{eq:time_adaptation}
\TimeEst < \mu \, \tola.
\end{equation}
where $\tola$ is the user-defined threshold tolerance and $\mu$ is a safety factor, which we take here equal to $\sqrt{10}$.
Accordingly, if condition \eqref{eq:time_adaptation} is met, the solution is accepted, 
otherwise, the solution is rejected and a new solution $\uVec_h|_{t+\delta t}, p_h|_{t+\delta t}$ is computed
restarting the time marching strategy from the previous step solution $\uVec_h|_{t}, p_h|_{t}$ and reducing the time step.

Several strategies for tailoring the time step are available in the literature, see \cite{OrdinaryII,LangROS3PL,Soderlind:2003,
Noventa.Massa:2020, Ghidoni.noventa:2022} for some examples.
In this work, the time step update $\delta t_{u}$ is computed as follows
\begin{equation}\label{eq:dt_update}
   \delta t_{u} = \left(\dfrac{\tola}{\TimeEst}\right)^{\frac{1}{q}} \delta t,
\end{equation}
where $q$ is the order of accuracy of the error estimator $\TimeEst$, which matches the order of accuracy of the embedded solution. 
Notice that a time step reduction is guaranteed if the solution is rejected, \ie $\TimeEst> \mu \tola$.
Moreover, in order to prevent excessive time step fluctuations, the following limiter function is introduced
\begin{equation}\label{eq:dt_limiting}
   \delta t_{l}
   		=
   			\delta t
   			\left[
   			1  +\kappa\,\text{atan}\left(\dfrac{\delta t_{u} -\delta t}{\kappa \delta t}\right)
   			\right],
\end{equation}
where the free parameter $\kappa$ controls the limiting strategy.
In this work, $\kappa$ is chosen such that the maximum acceptable time step relaxation
ensures a one order of magnitude increase in the time integration error, \ie
\begin{equation*}
	\kappa	=	\dfrac{2}{\pi} \left( 10^{\frac{1}{q}} -1\right).
\end{equation*}
Upon completion of a time step, regardless of whether the solution is accepted or rejected, the new time step is computed
appling \eqref{eq:dt_update} and \eqref{eq:dt_limiting}, and, thus, setting $\delta t = \delta t_{l}$.

\section{Numerical validation}\label{sec:numerical.validation}
Let us briefly describe the numerical setup common to the test cases presented in what follows.
Numerical integration over mesh cells and mesh faces is performed over standardized reference cells based on Gaussian quadrature rules.
Reference-to-physical-frame mappings from reference entities to mesh entities are defined by means of Lagrange polynomials.
Discrete polynomial spaces over mesh cells and mesh faces are spanned by orthonormal modal bases.
Specifically, we rely on physical frame polynomial spaces over mesh cells
and reference frame polynomial spaces over mesh faces, as proposed in \cite{Botti.Di-Pietro:18}.
Orthogonalization in the physical frame is performed cell-by-cell starting from a
monomial basis defined in a local reference frame aligned with the principal axes of inertia of each mesh cell, as described in \cite{Bassi.Botti.ea:12}.

Simultaneous multi-process execution of the code is achieved partitioning the computational grid,
such that the number of mesh partitions matches the number of processes, 
and relying on the Message Passing Interface (MPI) to synchronize and exchange data among the processes.
Scalability of the iterative solvers applied for the resolution of algebraic problems is pursued relying on the PETSc library \cite{petscTR}.
In particular, we apply one sweep of $p$-multigrid V-cycle as a preconditioner for the FGMRES (fexible GMRES) iteration.
As a smoothing strategy within the V-cycle we rely on a few iterations of GMRES and 
we employ additive Schwarz method (ASM) preconditioners to further improve performance in parallel.
All applications of prolongation and restriction operators involved in the V-cycle
iteration are performed matrix-free, that is, without assembling the global sparse matrices
associated to the operators, see \cite{Botti.Di-Pietro:22} for details.

\subsection{Travelling waves} \label{sec:TWt}

In order to numerically assess the spatial and temporal convergence rates of the ESDIRK-HHO formulation, we consider the following 2D dimensionless damped travelling waves solution of the Navier--Stokes equations:
\begin{align*}
    u(x,y,t)  &=  1 + 2\cos(2\pi (x-t))\sin(2\pi (y-t)) e^{-8\pi^2\nu t}, \\
    v(x,y,t)  &=  1 - 2\sin(2\pi (x-t))\cos(2\pi (y-t)) e^{-8\pi^2\nu t}, \\
    p(x,y,t)    &=  - [\cos(4\pi (x-t)) + \cos(4\pi (y-t))] e^{-16\pi^2\nu t},
\end{align*}
where $u$ and $v$ are the velocity components in the $x$ and $y$ directions, respectively, $p$ is the pressure and $t$ is the time.
The computational domain is the unit square $\Omega = [0.25,1.25]\times[0.5,1.5]$ and time integration is carried out in the interval $[0,1]$.
Accordingly the errors in $L^2$ norm for the velocity, the velocity gradients and the pressure are computed based on the exact solution at the final time $\tF = 1$.
Periodic boundary conditions are enforced on $\partial \Omega$ and the velocity field is initialized interpolating the exact solution.
We rely on triangular cells meshes obtained subdividing in two right triangles the square cells of $N^2$ Cartesian grids, with $N$ cells per Cartesian direction.
Accordingly, the mesh cardinality is $\card{\Th}=2\,N^2$ when $h = \frac{\sqrt{2}}{N}$.

The temporal accuracy is tested with $\nu = \Reynolds^{-1} = 10^{-2}$ based on a $k=9$ HHO formulation on a triangular cells mesh with $N=8$ and $\card{\Th}=128$.
The results are reported in Table \ref{tab:TW_TConv} considering $\delta t = \frac{0.1}{2^i}$, $i = 0,\dots,4$, and ESDIRK schemes with third, fourth and fifth order of accuracy.
The numerical convergence rates clearly replicate the expected rates of error reduction.
It is interesting to notice that the pressure error in the $L^2$-norm stagnates around a precision of $10^{-9}$.
We remark that, due to the use of periodic boundary conditions,
the pressure mean value needs to be pinned by means of the Lagrange multipliers method.
Also the pressure gradient error in $L^2$ norm is reported.

The spatial accuracy is tested varying the viscosity, up to the inviscid limit.
In particular, we set $\nu = 10^{-1,-3,-6,-13}$ in order to cover both the viscous dominated and the convection-dominated flow regimes.
The results are reported in Tables 
\ref{tab:TW_SConvP0}--\ref{tab:TW_SConvP4}
for $k=0,1,2,3,4$ HHO formulations, respectively, considering five grids of a triangular mesh sequence with $N=4 \times {2^i}$, $i = 0,\dots,5$.
Time integration is performed relying on a fifth order accurate ESDIRK scheme with time step $\delta t = 1/400$.
A $k+1$ rate of error reduction is observed for the pressure error in the $L^2$ norm, both in the diffusion-dominated and convection-dominated regimes.
{
On the other hand, the velocity and velocity gradient errors in $L^2$ norm show rates of convergence of $k+2$ and $k+1$, respectively, in the diffusion-dominated regime,
whereas the rate of error reduction is reduced by half an order in the convection-dominated regime, hence achieving rates of convergence of $k+\frac{3}{2}$ and $k+\frac{1}{2}$, respectively.
We remark that the numerical convergence rates observed for the travelling waves test case match with the theoretical convergence rates estimates of \cite{Beirao-da-Veiga.Di-Pietro.ea:25}.}

\begin{table}
\centering
\small
\begin{tabular}{ccccccccccc}
$\delta t$         & $\| \nabla \uVec_h \|_{L^2(\Omega)}$ & rate &     $\| \uVec_h \|_{L^2(\Omega)}$   & rate   & $\| \nabla p_h \|_{L^2(\Omega)}$ & rate &  $\| p_h \|_{L^2(\Omega)}$  & rate   &  $\| \nabla \cdot \uVec_h \|_{L^2(\Omega)}$ \\
\toprule
0.1          & 1.5621      & --         & 0.1758    & --         & 1.0031     & --        & 0.07982   & --         & 2.26e-12    \\
0.05         & 0.2832     & 2.46        & 0.03187   & 2.46       & 0.1818    & 2.46       & 0.01447   & 2.46       & 1.60e-12    \\
0.025        & 0.03938    & 2.85        & 0.004432   & 2.85      & 0.02528   & 2.85       & 0.002012   & 2.85       & 6.08e-13    \\
0.0125       & 0.005063    & 2.96       & 0.0005698   & 2.96     & 0.003251  & 2.96       & 0.0002587   & 2.96       & 2.60e-14    \\
0.00625      & 0.0006381    & 2.99      & 7.1820e-05 & 2.99      & 0.0004097  & 2.99      & 3.2609e-05 & 2.99       & 3.30e-15    \\
\toprule
0.1          & 0.10181     & --         & 0.01145   & --         & 0.06537   & --         & 0.005202  & --         & 4.83e-12  \\
0.05         & 0.006672   & 3.93        & 0.0007509   & 3.93     & 0.004284  & 3.93       & 0.0003409  & 3.93       & 1.20e-12  \\
0.025        & 0.0004210   & 3.99       & 4.7386e-05 & 3.99      & 0.0002703  & 3.99      & 2.1515e-05 & 3.99       & 6.54e-14  \\
0.0125       & 2.6370e-05  & 4.00       & 2.9677e-06 & 4.00      & 1.6932e-05 & 4.00      & 1.3474e-06 & 4.00       & 1.57e-13  \\
0.00625      & 1.6488e-06  & 4.00       & 1.8555e-07 & 4.00      & 1.0586e-06 & 4.00      & 8.4531e-08 & 3.99       & 3.03e-14  \\
\toprule
0.1          & 0.01553    & --         & 0.001748   & --         & 0.009977  & --         & 0.0007939  & --         & 6.46e-12       \\
0.05         & 0.0005485   & 4.82       & 6.1735e-05 & 4.82      & 0.0003522  & 4.82      & 2.8030e-05 & 4.82       & 1.51e-12       \\
0.025        & 1.7765e-05  & 4.95       & 1.9992e-06 & 4.95      & 1.1407e-05 & 4.95      & 9.0776e-07 & 4.95       & 4.32e-13       \\
0.0125       & 5.6137e-07  & 4.98       & 6.3176e-08 & 4.98      & 3.6062e-07 & 4.98      & 2.8768e-08 & 4.98       & 4.33e-13        \\
0.00625      & 1.7691e-08  & 4.99       & 1.9818e-09 & 4.99      & 1.1676e-08 & 4.95      & 9.551e-09   & 1.59       & 2.36e-13       \\
\bottomrule                                                                                                                                                                     \end{tabular}
\caption{Numerical validation of third, fourth and fifth order accurate ESDIRK time integration strategies (first, second and third bunch of rows, respectively) based on the travelling waves analytical solution, $\nu=0.01$.
$L^2(\Omega)$ error norms with respect to the exact solution and the corresponding rates of convergence while halving the time step $\delta t$ are reported.
The spatial discretization is chosen such that the spatial approximation error is negligible, see text for details. \label{tab:TW_TConv}}
\end{table}

\begin{table}
\small
\centering
\begin{tabular}{ccccccccccc}
$\nu$         & $\card{\Th}$ & $\| \nabla \uVec_h \|_{L^2(\Omega)}$ & rate &     $\| \uVec_h \|_{L^2(\Omega)}$   & rate
                          & $\| \nabla p_h \|_{L^2(\Omega)}$ & rate &     $\| p_h \|_{L^2(\Omega)}$  & rate   &  $\| \nabla \cdot \uVec_h \|_{L^2(\Omega)}$ \\
0.1        & 128        & 0.01148    &  --     & 0.001273 &   --     & --    & --      & 2.0701e-05 &   --    & 1.67e-15    \\ 
0.001      & 128        & 10.1662    &  --     & 1.0338   &   --     & --    & --      & 0.9074     &   --     & 7.51e-15    \\ 
1e-06      & 128        & 11.1946    &  --     & 1.1038   &   --     & --    & --      & 1.0468     &   --     & 1.04e-14    \\ 
1e-13      & 128        & 11.1958    &  --     & 1.1038   &   --     & --    & --      & 1.0469     &   --     & 9.17e-15    \\ 
\toprule
0.1        & 512        & 0.002402 & 2.26       & 0.0002397 & 2.41       & --    & --      & 2.3294e-06 & 3.15       & 1.03e-15       \\ 
0.001      & 512        & 6.4406   & 0.66       & 0.6161   & 0.75        & --    & --      & 0.5744     & 0.66       & 8.44e-15       \\ 
1e-06      & 512        & 6.6716   & 0.75       & 0.6476   & 0.77        & --    & --      & 0.6601     & 0.67       & 7.89e-15       \\ 
1e-13      & 512        & 6.6748   & 0.75       & 0.6476   & 0.77        & --    & --      & 0.6602     & 0.67       & 7.96e-15       \\ 
\toprule
0.1        & 2048       & 0.0006288 & 1.93       & 5.1296e-05 & 2.22     &  --    & --     & 8.0987e-07 & 1.52     & 1.10e-15     \\ 
0.001      & 2048       & 5.8408    & 0.14       & 0.3049     & 1.01     &  --    & --     & 0.2898   & 0.99       & 9.39e-15     \\ 
1e-06      & 2048       & 3.4654    & 0.95       & 0.3102     & 1.06     &  --    & --     & 0.3178   & 1.05       & 8.79e-15     \\ 
1e-13      & 2048       & 3.4819    & 0.94       & 0.3101     & 1.06     &  --    & --     & 0.3177   & 1.06       & 9.03e-15     \\ 
\toprule
0.1        & 8192       & 0.0002298 & 1.45       & 1.2142e-05 & 2.08     &  -- & --      & 3.8258e-07 & 1.08     & 3.42e-16    \\ 
0.001      & 8192       & 2.5467    & 1.20       & 0.08071    & 1.92     &  -- & --      & 0.08423  & 1.78       & 9.59e-15    \\ 
1e-06      & 8192       & 1.6416    & 1.08       & 0.1362     & 1.19     &  -- & --      & 0.1437   & 1.14       & 9.61e-15    \\ 
1e-13      & 8192       & 1.6948    & 1.04       & 0.1359     & 1.19     &  -- & --      & 0.1434   & 1.15       & 9.29e-15    \\ 
\toprule
0.1        & 32768      & 0.0001033 & 1.15      & 2.9733e-06 & 2.03     &    -- & --     & 1.9088e-07 & 1.00     & 3.68e-15   \\ 
0.001      & 32768      & 0.6611   & 1.95       & 0.01257  & 2.68       &    -- & --     & 0.02282  & 1.88       & 9.27e-15   \\ 
1e-06      & 32768      & 1.1604    & 0.50      & 0.06522  & 1.06       &    -- & --     & 0.07040  & 1.03       & 1.00e-14   \\ 
1e-13      & 32768      & 0.9233   & 0.88       & 0.06474  & 1.07       &    -- & --     & 0.06971  & 1.04       & 9.59e-15   \\ 
\bottomrule
\end{tabular}
\caption{Numerical validation of spatial convergence of $k=0$ HHO formulations based on the 2D travelling waves analytical solution of the INS equations.
$L^2(\Omega)$ error norms with respect to the exact solution and the corresponding rates of convergence while halving the mesh step size are reported.
The temporal discretization is chosen such that the temporal approximation error is negligible, see text for details. \label{tab:TW_SConvP0}}
\end{table}

\begin{table}
\small
\centering
\begin{tabular}{ccccccccccc}
$\nu$         & $\card{\Th}$ & $\| \nabla \uVec_h \|_{L^2(\Omega)}$ & rate &     $\| \uVec_h \|_{L^2(\Omega)}$   & rate
                          & $\| \nabla p_h \|_{L^2(\Omega)}$ & rate &     $\| p_h \|_{L^2(\Omega)}$  & rate   &  $\| \nabla \cdot \uVec_h \|_{L^2(\Omega)}$ \\
\toprule
0.1        & 128        & 0.000409  &  --        & 2.1611e-05 &   --     & 0.0001346   &   --    & 9.2621e-06 &   --    & 3.74e-14       \\ 
0.001      & 128        & 2.3374    &  --       & 0.1179      &   --       & 4.1976    &   --      & 0.1202   &   --      & 4.18e-12       \\ 
1e-06      & 128        & 2.7412    &  --       & 0.1406      &   --       & 4.9579    &   --      & 0.1499   &   --      & 6.42e-14       \\ 
1e-13      & 128        & 2.7420    &  --       & 0.1407      &   --       & 4.9587    &   --      & 0.1500   &   --      & 6.50e-14       \\ 
\toprule
0.1        & 512        & 7.5701e-05 & 2.43     & 1.0458e-06 & 4.37     & 3.0704e-05 & 2.13      & 7.6158e-06 & 0.28     & 9.84e-14       \\ 
0.001      & 512        & 0.8812   & 1.41       & 0.01157   & 3.35      & 2.0514    & 1.03       & 0.01656  & 2.86       & 1.19e-12       \\ 
1e-06      & 512        & 0.8271    & 1.73      & 0.01439  & 3.29       & 2.4063    & 1.04       & 0.02121   & 2.82      & 1.95e-11       \\ 
1e-13      & 512        & 0.8273   & 1.73       & 0.01439  & 3.29       & 2.4067    & 1.04       & 0.02121  & 2.82       & 1.93e-11       \\ 
\toprule
0.1        & 2048       & 1.8367e-05 & 2.04     & 7.5911e-08 & 3.78     & 8.9306e-06 & 1.78      & 5.8233e-07 & 3.71     & 1.46e-13       \\ 
0.001      & 2048       & 0.1632   & 2.43       & 0.000717 & 4.01       & 1.0303     & 0.99      & 0.003807 & 2.12       & 9.88e-13       \\ 
1e-06      & 2048       & 0.2324   & 1.83       & 0.001311 & 3.46       & 1.2065    & 1.00       & 0.004485 & 2.24       & 1.21e-12       \\ 
1e-13      & 2048       & 0.2322   & 1.83       & 0.001308 & 3.46       & 1.2067    & 1.00       & 0.004486  & 2.24      & 1.20e-12       \\ 
\toprule
0.1        & 8192       & 4.5830e-06 & 2.00     & 7.8398e-09 & 3.28     & 3.3104e-06 & 1.43      & 2.5148e-08 & 4.53     & 2.69e-13    \\ 
0.001      & 8192       & 0.02495  & 2.71       & 5.4704e-05 & 3.71     & 0.51586   & 1.00       & 0.0009505 & 2.00      & 6.47e-12    \\ 
1e-06      & 8192       & 0.06771  & 1.78       & 0.0001702 & 2.95      & 0.60401   & 1.00       & 0.001113 & 2.01       & 8.42e-12    \\ 
1e-13      & 8192       & 0.06733  & 1.79       & 0.0001687 & 2.96      & 0.60411   & 1.00       & 0.001113 & 2.01       & 6.05e-12    \\ 
\toprule
0.1        & 32768      & 1.1469e-06 & 2.00     & 9.3604e-10 & 3.07       & 1.4688e-06 & 1.17    & 3.2693e-09 & 2.13      & 3.16e-14  \\ 
0.001      & 32768      & 0.004410 & 2.50       & 4.4184e-06 & 3.63       & 0.2580   & 1.00      & 0.0002375 & 2.00       & 3.69e-13  \\ 
1e-06      & 32768      & 0.02064  & 1.71       & 2.4545e-05 & 2.79       & 0.3021    & 1.00     & 0.0002782 & 2.00       & 3.27e-13  \\ 
1e-13      & 32768      & 0.01980  & 1.77       & 2.3734e-05 & 2.83       & 0.3021   & 1.00      & 0.0002782 & 2.00       & 3.16e-13  \\ 
\bottomrule
\end{tabular}
\caption{Numerical validation of spatial convergence of $k=1$ HHO formulations based on the 2D travelling waves analytical solution of the INS equations.
$L^2(\Omega)$ error norms with respect to the exact solution and the corresponding rates of convergence while halving the mesh step size are reported.
The temporal discretization is chosen such that the temporal approximation error is negligible, see text for details. \label{tab:TW_SConvP1}}
\end{table}

\begin{table}
\centering
\small
\begin{tabular}{ccccccccccc}
$\nu$         & $\card{\Th}$ & $\| \nabla \uVec_h \|_{L^2(\Omega)}$ & rate &     $\| \uVec_h \|_{L^2(\Omega)}$   & rate
                          & $\| \nabla p_h \|_{L^2(\Omega)}$ & rate &     $\| p_h \|_{L^2(\Omega)}$  & rate   &  $\| \nabla \cdot \uVec_h \|_{L^2(\Omega)}$ \\
\toprule
0.1        & 32         & 0.0006025 & --        & 3.2519e-05 & --       & 0.0002892 & --      & 9.0898e-06 & --       & 1.65e-15     \\ 
0.001      & 32         & 2.1575    & --        & 0.09630  & --         & 3.95983    & --         & 0.1024    & --        & 1.30e-14     \\ 
1e-06      & 32         & 2.4141    & --        & 0.1114   & --         & 4.64252    & --         & 0.1194   & --         & 1.35e-14     \\ 
1e-13      & 32         & 2.4144    & --        & 0.1114    & --        & 4.64326    & --        & 0.1194   & --         & 1.28e-14     \\ 
\toprule
0.1        & 128        & 4.8376e-05 & 3.64     & 5.0157e-07 & 6.02     & 3.8012e-05 & 2.93      & 6.9970e-07 & 3.70      & 1.67e-15     \\ 
0.001      & 128        & 0.3370   & 2.68       & 0.007253 & 3.73       & 0.82363   & 2.27       & 0.008592 & 3.58        & 3.97e-14     \\ 
1e-06      & 128        & 0.4121    & 2.55      & 0.009759 & 3.51       & 0.96524    & 2.27      & 0.01061   & 3.49       & 9.39e-14     \\ 
1e-13      & 128        & 0.4123   & 2.55       & 0.009763 & 3.51       & 0.96539   & 2.27       & 0.01061  & 3.49        & 9.36e-14      \\ 
\toprule
0.1        & 512        & 5.8504e-06 & 3.05     & 2.2680e-08 & 4.47      & 7.6681e-06 & 2.31     & 8.4712e-08 & 3.05      & 2.70e-15     \\ 
0.001      & 512        & 0.08293  & 2.02       & 0.0004579 & 3.99       & 0.2100   & 1.97       & 0.0009889 & 3.12       & 3.34e-14     \\ 
1e-06      & 512        & 0.07475  & 2.46       & 0.0008462 & 3.53       & 0.2459   & 1.97       & 0.001171 & 3.18        & 4.15e-14     \\ 
1e-13      & 512        & 0.07491  & 2.46       & 0.0008479 & 3.53       & 0.2460   & 1.97       & 0.001171 & 3.18        & 4.14e-14     \\ 
\toprule
0.1        & 2048       & 7.364e-07  & 2.99     & 1.3936e-09 & 4.02       & 1.8567e-06 & 2.05     & 1.1375e-08 & 2.90     & 2.78e-15     \\ 
0.001      & 2048       & 0.01009  & 3.04       & 2.3197e-05 & 4.30       & 0.05277   & 1.99      & 0.0001230 & 3.01      & 1.41e-13     \\ 
1e-06      & 2048       & 0.01594  & 2.23       & 9.7367e-05 & 3.12       & 0.06179  & 1.99       & 0.0001442 & 3.02      & 2.29e-13     \\ 
1e-13      & 2048       & 0.01610  & 2.22       & 9.8386e-05 & 3.11       & 0.06180  & 1.99       & 0.0001442 & 3.02      & 2.30e-13     \\ 
\toprule
0.1        & 8192       & 9.2628e-08 & 2.99     & 8.7775e-11 & 3.99       & 4.6574e-07 & 2.00     & 1.4261e-09 & 3.00     & 5.22e-15     \\ 
0.001      & 8192       & 0.0006961 & 3.86      & 8.0790e-07 & 4.84       & 0.01320  & 2.00       & 1.5388e-05 & 3.00     & 3.98e-14     \\ 
1e-06      & 8192       & 0.002882 & 2.47       & 9.0767e-06 & 3.42       & 0.01546  & 2.00       & 1.8022e-05 & 3.00     & 6.82e-14     \\ 
1e-13      & 8192       & 0.002854 & 2.50       & 9.3780e-06 & 3.39       & 0.01546  & 2.00       & 1.8025e-05 & 3.00     & 6.83e-14     \\ 
\bottomrule
\end{tabular}
\caption{Numerical validation of spatial convergence of $k=2$ HHO formulations based on the 2D travelling waves analytical solution of the INS equations.
$L^2(\Omega)$ error norms with respect to the exact solution and the corresponding rates of convergence while halving the mesh step size are reported.
The temporal discretization is chosen such that the temporal approximation error is negligible, see text for details. \label{tab:TW_SConvP2}}
\end{table}

\begin{table}
\centering
\small
\begin{tabular}{ccccccccccc}
$\nu$         & $\card{\Th}$ & $\| \nabla \uVec_h \|_{L^2(\Omega)}$ & rate &     $\| \uVec_h \|_{L^2(\Omega)}$   & rate
                          & $\| \nabla p_h \|_{L^2(\Omega)}$ & rate &     $\| p_h \|_{L^2(\Omega)}$  & rate   &  $\| \nabla \cdot \uVec_h \|_{L^2(\Omega)}$ \\
\toprule
0.1        & 32         & 0.0001296 & --        & 2.1174e-06 & --       & 6.1878e-05 & --       & 1.5895e-06 & --       & 1.03e-15     \\ 
0.001      & 32         & 0.4937   & --         & 0.01724  & --         & 0.3720    & --        & 0.01391  & --         & 8.17e-15     \\ 
1e-06      & 32         & 0.7047   & --         & 0.02619  & --         & 0.4843   & --         & 0.02122  & --         & 2.78e-14     \\ 
1e-13      & 32         & 0.7050   & --         & 0.02621  & --         & 0.4845     & --       & 0.02123  & --         & 2.76e-14     \\ 
\toprule
0.1        & 128        & 5.3691e-06 & 4.59     & 3.0734e-08 & 6.11      & 5.1743e-06 & 3.58    & 5.4149e-08 & 4.89    & 5.97e-13     \\ 
0.001      & 128        & 0.04189  & 3.56       & 0.0004808 & 5.16       & 0.1142   & 1.70      & 0.0007852 & 4.15     & 1.38e-14     \\ 
1e-06      & 128        & 0.05105  & 3.79       & 0.0007212 & 5.18       & 0.1341   & 1.85      & 0.0009479 & 4.48     & 4.08e-14     \\ 
1e-13      & 128        & 0.05108  & 3.79       & 0.0007216 & 5.18       & 0.1341   & 1.85      & 0.0009481 & 4.48     & 4.20e-14     \\ 
\toprule
0.1        & 512        & 3.1562e-07 & 4.09     & 8.6284e-10 & 5.15       & 6.6143e-07 & 2.97   & 3.6144e-08 & 3.90     & 6.91e-14     \\ 
0.001      & 512        & 0.003648 & 3.52       & 1.3916e-05 & 5.11       & 0.01449  & 2.98     & 4.7942e-05 & 4.03     & 2.75e-13     \\ 
1e-06      & 512        & 0.004017 & 3.67       & 2.4970e-05 & 4.85       & 0.01697  & 2.98     & 5.6211e-05 & 4.08     & 5.48e-14     \\ 
1e-13      & 512        & 0.004019 & 3.67       & 2.4997e-05 & 4.85       & 0.01698  & 2.98     & 5.6220e-05 & 4.08     & 5.36e-14      \\ 
\toprule
0.1        & 2048       & 1.9790e-08 & 4.00       & 2.6989e-11 & 5.00       & 8.4707e-08 & 2.97  & 7.3114e-09 & -1.02   & 1.56e-12      \\ 
0.001      & 2048       & 0.0001470 & 4.63       & 2.4832e-07 & 5.81       & 0.001820 & 2.99    & 3.0008e-06 & 4.00     & 2.34e-13      \\ 
1e-06      & 2048       & 0.0002740 & 3.87       & 8.7129e-07 & 4.84       & 0.002131 & 2.99    & 3.5143e-06 & 4.00     & 1.65e-13      \\ 
1e-13      & 2048       & 0.0002736 & 3.88       & 8.7385e-07 & 4.84       & 0.002131  & 2.99   & 3.5148e-06 & 4.00     & 1.56e-13      \\ 
\toprule
0.1        & 8192       & 1.2450e-09 & 3.99       & 8.5138e-13 & 4.99       & 1.0713e-08 & 2.98       & 9.9058e-10 & 2.88       & 7.01e-15     \\ 
0.001      & 8192       & 6.2375e-06 & 4.56       & 1.0842e-08 & 4.52       & 0.000227836 & 3.00       & 1.87872e-07 & 4.00    & 2.57e-14      \\ 
1e-06      & 8192       & 1.9812e-05 & 3.79       & 3.3310e-08 & 4.71       & 0.000266768 & 3.00       & 2.20079e-07 & 4.00    & 1.86e-13      \\ 
1e-13      & 8192       & 1.9625e-05 & 3.80       & 3.3607e-08 & 4.70       & 0.000266809 & 3.00       & 2.20115e-07 & 4.00     & 1.94e-13       \\ 
\bottomrule
\end{tabular}
\caption{Numerical validation of spatial convergence of $k=3$ HHO formulations based on the 2D travelling waves analytical solution of the INS equations.
$L^2(\Omega)$ error norms with respect to the exact solution and the corresponding rates of convergence while halving the mesh step size are reported.
The temporal discretization is chosen such that the temporal approximation error is negligible, see text for details. \label{tab:TW_SConvP3}}
\end{table}
 
\begin{table}
\centering
\small
\begin{tabular}{ccccccccccc}
$\nu$         & $\card{\Th}$ & $\| \nabla \uVec_h \|_{L^2(\Omega)}$ & rate &     $\| \uVec_h \|_{L^2(\Omega)}$   & rate
                          & $\| \nabla p_h \|_{L^2(\Omega)}$ & rate &     $\| p_h \|_{L^2(\Omega)}$  & rate   &  $\| \nabla \cdot \uVec_h \|_{L^2(\Omega)}$ \\
\toprule
0.1        & 32         & 2.0021e-05 & --       & 2.0315e-07 & --       & 1.4847e-05 & --         & 2.2925e-07 & --       & 1.14e-13   \\ 
0.001      & 32         & 0.09759  & --         & 0.001717 & --         & 0.250568   & --         & 0.002552 & --         & 3.33e-14   \\ 
1e-06      & 32         & 0.1364   & --         & 0.002667 & --         & 0.294825   & --         & 0.003198 & --         & 3.50e-14   \\ 
1e-13      & 32         & 0.1365   & --         & 0.002669 & --         & 0.294876   & --         & 0.003199 & --         & 3.32e-14   \\ 
\toprule
0.1        & 128        & 4.8000e-07 & 5.38     & 2.0321e-09 & 6.64     & 5.6606e-07 & 4.71     & 5.9306e-08 & 1.95       & 1.81e-12   \\ 
0.001      & 128        & 0.003536 & 4.79       & 2.9056e-05 & 5.89     & 0.01183  & 4.40       & 5.9769e-05 & 5.42       & 5.77e-14   \\ 
1e-06      & 128        & 0.005113 & 4.74       & 5.7780e-05 & 5.53     & 0.01385  & 4.41       & 7.1877e-05 & 5.48       & 2.01e-13   \\ 
1e-13      & 128        & 0.005118 & 4.74       & 5.7824e-05 & 5.53     & 0.01385  & 4.41       & 7.1897e-05 & 5.48       & 2.06e-13   \\ 
\toprule
0.1        & 512        & 1.4710e-08 & 5.03      & 2.9976e-11 & 6.08    & 3.1149e-08 & 4.18      & 1.1526e-08 & 2.36       & 4.50e-13   \\ 
0.001      & 512        & 0.0002101 & 4.07       & 5.1621e-07 & 5.81    & 0.0007507 & 3.98       & 1.8745e-06 & 4.99       & 4.02e-13   \\ 
1e-06      & 512        & 0.0002313 & 4.47       & 1.2782e-06 & 5.50    & 0.0008790 & 3.98       & 2.2159e-06 & 5.02       & 6.16e-14   \\ 
1e-13      & 512        & 0.0002322 & 4.46       & 1.2827e-06 & 5.49    & 0.0008791 & 3.98       & 2.2165e-06 & 5.02       & 4.17e-14   \\ 
\toprule
0.1        & 2048       & 4.6501e-10 & 4.98       & 4.7288e-13 & 5.99   & 1.9160e-09 & 4.02       & 1.3833e-09 & 3.06       & 7.93e-15     \\ 
0.001      & 2048       & 5.3305e-06 & 5.30       & 6.1381e-09 & 6.39   & 4.7104e-05 & 3.99       & 6.2906e-08 & 4.90       & 3.00e-13     \\ 
1e-06      & 2048       & 9.9244e-06 & 4.54       & 3.0864e-08 & 5.37   & 5.5154e-05 & 3.99       & 6.8994e-08 & 5.01       & 3.85e-13     \\ 
1e-13      & 2048       & 1.0025e-05 & 4.53       & 3.1290e-08 & 5.36   & 5.5163e-05 & 3.99       & 6.8999e-08 & 5.01       & 5.50e-13     \\ 
\toprule
0.1        & 8192       & 1.5221e-11 & 4.93       & 4.5712e-14 & 3.37   & 1.8249e-10 & 3.39       & 8.5895e-09 & -2.63      & 2.16e-15     \\ 
0.001      & 8192       & 9.3105e-08 & 5.84       & 6.9149e-11 & 6.47   & 2.9470e-06 & 4.00       & 2.0341e-09 & 4.95       & 2.94e-13     \\ 
1e-06      & 8192       & 4.3592e-07 & 4.51       & 6.8307e-10 & 5.50   & 3.4506e-06 & 4.00       & 2.2686e-09 & 4.93       & 5.79e-13     \\ 
1e-13      & 8192       & 3.8003e-07 & 4.72       & 7.0825e-10 & 5.47   & 3.4511e-06 & 4.00       & 2.2817e-09 & 4.92       & 3.17e-13     \\ 
\bottomrule
\end{tabular}
\caption{Numerical validation of spatial convergence rates of $k=4$ HHO formulations, 2D travelling waves analytical solution of the INS equations.
$L^2(\Omega)$ error norms with respect to the exact solution and the corresponding rates of convergence while halving the mesh step size are reported.
The temporal discretization is chosen such that the temporal approximation error is negligible, see text for details. \label{tab:TW_SConvP4}}
\end{table}

\subsection{Double shear layer}\label{sec:DSLayer}

The double shear layer problem, introduced in \cite{Bell.et.al:1989},
is employed to assess the ESDIRK-HHO formulation in the context of unsteady inviscid flows (\ie $\nu=0$).
The computational domain is double periodic unit square, $\Omega=\left(0,\,1\right)\times\left(0,\,1\right)$ and time
integration is carried out over the time interval $[t_0 = 0, \tF = 2]$.
The initial conditions for the horizontal and vertical velocity components read
\begin{equation*}
    u   =   \begin{cases}
                \tanh{\left(\dfrac{y-0.25}{\xi}\right)}    &   \text{if $y\leq 0.5$}, \\
                \tanh{\left(\dfrac{0.75-y}{\xi}\right)}    &   \text{if $y> 0.5$}
            \end{cases}
    \quad \text{and} \quad v   =   \delta \sin{\rbrackets{2\pi x}},
\end{equation*}
respectively.
Free parameters are set as $\xi   =   1/30$ and $\delta  =   1/20$.

Since an inviscid flow is non-dissipative and no power is introduced in $\Omega$ 
from the sorroundings (basically, due to the imposition of periodic 
boundary conditions, there are no sourroundings), the time evolution of the average kinetic energy over $\Omega$ 
quantifies the numerical dissipation introduced by the numerical scheme.
In particular, we evaluate numerical dissipation based on the normalized kinetic energy error 
computed as follows: $(\mathcal{K}_0 - \mathcal{K}_F) / \mathcal{K}_0$, 
where $\mathcal{K}_{0}=\int_\Omega \frac{1}{2}\uVec_h|_{t_0} \cdot \uVec_h|_{t_0}$ is the initial kinetic energy
and $\mathcal{K}_{\rm F}= \int_\Omega \frac{1}{2} \uVec_h|_{\tF} \cdot \uVec_h|_{\tF}$ is the final kinetic energy.

We consider $k=1,2,3,4$ HHO formulations on a
$h$-refined regular triangular mesh sequence, the mesh cardinality reads $\card{\Th} = 32 \times 4^i, \, i=0,\dots,6$.
Time integration is performed by means of a fifth-order accurate ESDIRK scheme with local time step adaptation.
The trigger tolerance for time step adaptation is chosen such that the temporal error contribution
weights less than $0.1\%$ of the overall relative kinetic energy error.
The tolerance and the estimated weight of the temporal error
are reported in Table~\ref{tab:DSL_adapt_tolerance} for $k=4$ HHO discretization and every grid in the mesh sequence.
Clearly, the highest polynomial degree is the most demanding in terms of time integration accuracy.
\begin{table}[t]
\centering
\small
\begin{tabular}{cccc}
$i$         & $\card{\Th}$ & $\tola$ & $\mathcal{I}_t^{\%}$ \\
\toprule
0  & 32         & 1e-06 & 0.02\%  \\
1  & 128        & 1e-06 & 0.04\%  \\
2  & 512        & 1e-07 & 0.04\%  \\
3  & 2048       & 1e-08 & 0.03\%  \\
4  & 8192       & 1e-09 & 0.04\%  \\
5  & 32768      & 1e-10 & 0.04\%  \\
6  & 131072     & 1e-11 & 0.05\%  \\
\bottomrule
\end{tabular}
\caption{Double shear layer, $k=4$ ESDIRK-HHO with local time step adaptation. Adaptation trigger tolerance ($\tola$) and
estimated relative weight of the temporal discretization error ($\mathcal{I}_t^{\%}$) over the kinetic energy error are reported.
\label{tab:DSL_adapt_tolerance}}
\end{table}

\begin{figure} [!t]
\centering
\begin{tikzpicture}[scale=0.65]
\begin{axis}    [
                unbounded coords=jump,
                ymode=log,xmode=log,
                scale only axis=true,
                width=0.5\textwidth,
                height=0.5\textwidth,
                xlabel=mesh spacing $h$,
                ylabel= $(\mathcal{K}_0 - \mathcal{K}_F) / \mathcal{K}_0$,
                label style={anchor=near ticklabel, font=\large},
                ticks=major,
                tick pos=left,
                tick align=center,
                xmin=3.125E-3,
                xmax=3.333E-1,
                ymin=2E-7,
                ymax=3E-1,
                legend style={cells={anchor=west}},
                legend pos=north west,
                ]
\addplot    [
            thick,
            mark = *,
            mark size=3pt,
            mark options={black, fill=white}
            ]
            table[
            x expr = {\thisrowno{0}},
            y expr = {\thisrowno{2}},
            header=false]
            {DSL_k_p1.dat};
\addplot    [
            thick,
            densely dashed,
            mark = triangle*,
            mark size=3pt,
            mark options={black, solid, fill=white}
            ]
            table[
            x expr = {\thisrowno{0}},
            y expr = {\thisrowno{2}},
            header=false]
            {DSL_k_p2.dat};
\addplot    [
            thick,
            dotted,
            mark = square*,
            mark size=3pt,
            mark options={black, solid, fill=white}
            ]
            table[
            x expr = {\thisrowno{0}},
            y expr = {\thisrowno{2}},
            header=false]
            {DSL_k_p3.dat};
\addplot    [
            thick,
            dashdotted,
            mark = diamond*,
            mark size=3pt,
            mark options={black, solid, fill=white}
            ]
            table[
            x expr = {\thisrowno{0}},
            y expr = {\thisrowno{2}},
            header=false]
            {DSL_k_p4.dat};
\legend {$k=1$, $k=2$, $k=3$, $k=4$}
\end{axis}
\end{tikzpicture}
\hspace{5mm}
\begin{tikzpicture}[scale=0.65]
\begin{axis}    [
                unbounded coords=jump,
                ymode=log,xmode=log,
                scale only axis=true,
                width=0.5\textwidth,
                height=0.5\textwidth,
                xlabel=Degrees of Freedom (DOFs),
                ylabel= $(\mathcal{K}_0 - \mathcal{K}_F) / \mathcal{K}_0$,
                label style={anchor=near ticklabel, font=\large},
                ticks=major,
                tick pos=left,
                tick align=center,
                xmin=2E+2,
                xmax=6E+6,
                ymin=2E-7,
                ymax=3E-1,
                legend style={cells={anchor=west}},
                legend pos=north east,
                ]
\addplot    [
            thick,
            mark = *,
            mark size=3pt,
            mark options={black, fill=white}
            ]
            table[
            x expr = {\thisrowno{1}},
            y expr = {\thisrowno{2}},
            header=false]
            {DSL_k_p1.dat};
\addplot    [
            thick,
            densely dashed,
            mark = triangle*,
            mark size=3pt,
            mark options={black, solid, fill=white}
            ]
            table[
            x expr = {\thisrowno{1}},
            y expr = {\thisrowno{2}},
            header=false]
            {DSL_k_p2.dat};
\addplot    [
            thick,
            dotted,
            mark = square*,
            mark size=3pt,
            mark options={black, solid, fill=white}
            ]
            table[
            x expr = {\thisrowno{1}},
            y expr = {\thisrowno{2}},
            header=false]
            {DSL_k_p3.dat};
\addplot    [
            thick,
            dashdotted,
            mark = diamond*,
            mark size=3pt,
            mark options={black, solid, fill=white}
            ]
            table[
            x expr = {\thisrowno{1}},
            y expr = {\thisrowno{2}},
            header=false]
            {DSL_k_p4.dat};
\end{axis}
\end{tikzpicture}
\caption{Double shear layer. \emph{Left and right}, relative kinetic energy error at the end time $\tF=2$.
as a function of the mesh spacing and of the total number of degrees of freedom, respectively. \label{fig:double_shear_layer_convergence_cost}}
\end{figure}
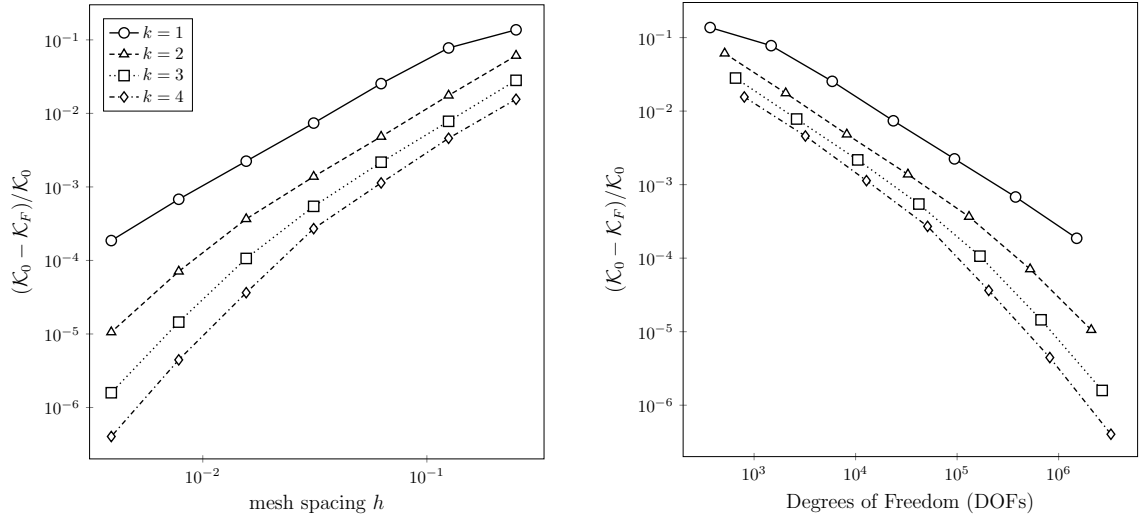

In Figure~\ref{fig:double_shear_layer_convergence_cost}, the relative kinetic energy error
is plotted against the mesh spacing or the total number degrees of freedom (DOFs).
As expected, finer meshes and higher polynomial degrees result in lower numerical dissipation.
Moreover, increasing the polynomial degree systematically enhances the rate of error reduction with respect to the number of DOFs.
Table~\ref{tab:DSL_convercgence_rates} reports the relative kinetic energy error along
with its rate of convergence for all grids and polynomial degrees combinations.
We remark that the asymptotic $k+3/2$ convergence rates are not fully achieved for $k=2,3,4$, 
only the $k=1$ HHO formulation shows the tendency to reach the expected rate of error reduction on the finest grids of the mesh sequence.

\begin{figure} [!t]
\centering
\includegraphics*[width=0.80\linewidth]{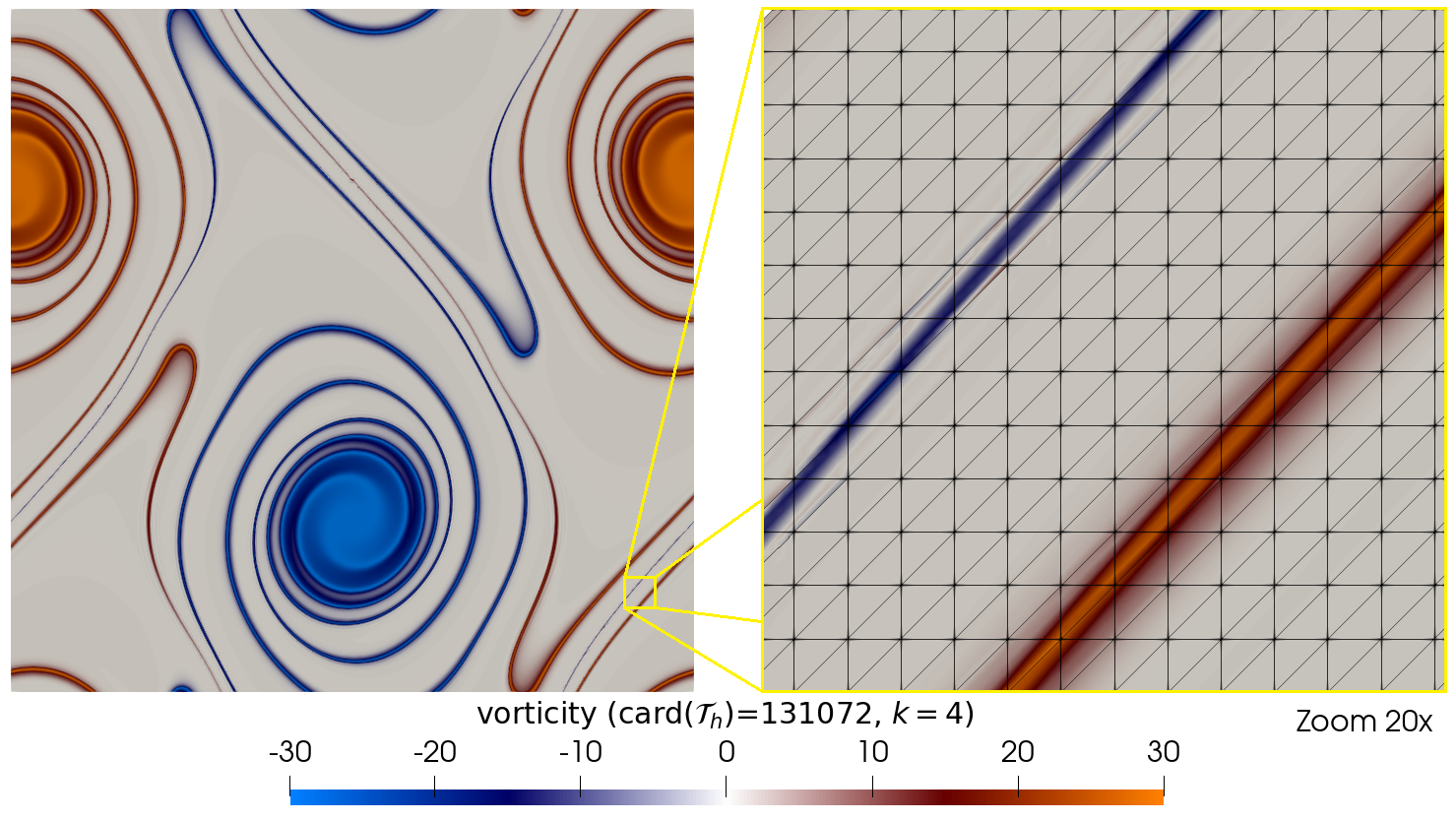}
\caption{Double shear layer. \emph{Left and right}: vorticity field at the end time $t = 2$ considering the whole computational domain
and a zoomed-in view with the computational grid superimposed.
The solution is obtained based on a $k = 4$ HHO formulation on the finest grid (left) relying on a fifth order accurate ESDIRK time marching strategy. \label{fig:double_shear_layer_vorticity}}
\end{figure}

It is quite evident that more pronounced discrepancies are observed for higher polynomial degrees.
Moreover, when the kinetic energy error
is higher than $10^{-4}$, the rate of error reduction does not exceed the second order, regardless of the polynomial degree.
\begin{table}
\centering
\small
\begin{tabular}{ccccccccccc}
$i$         & $\card{\Th}$ & error ($k=1$) & rate &     error ($k=2$)   & rate   & error ($k=3$) & rate &  error ($k=4$)  & rate \\
\toprule
0    & 32      &  1.263e-01  & --         & 6.228e-02   & --         & 3.166e-02  & --         & 1.821e-02 & --   \\
1    & 128     &  5.129e-02  & 1.30       & 1.876e-02   & 1.73       & 8.492e-03  & 1.90       & 4.735e-03 & 1.94 \\
2    & 512     &  1.984e-02  & 1.37       & 4.744e-03   & 1.98       & 2.047e-03  & 2.05       & 1.087e-03 & 2.12 \\
3    & 2048    &  5.337e-03  & 1.89       & 1.128e-03   & 2.07       & 4.793e-04  & 2.09       & 2.434e-04 & 2.16 \\
4    & 8192    &  1.494e-03  & 1.84       & 2.600e-04   & 2.12       & 8.191e-05  & 2.55       & 3.225e-05 & 2.92 \\
5    & 32768   &  4.177e-04  & 1.84       & 4.007e-05   & 2.70       & 1.080e-05  & 2.92       & 3.650e-06 & 3.14 \\
6    & 131072  &  8.906e-05  & 2.23       & 5.378e-06   & 2.90       & 1.165E-06  & 3.21       & 3.157e-07 & 3.53 \\
\bottomrule                                                                                                                                                                     \end{tabular}
\caption{Double shear layer. Numerical validation of spatial convergence of $k=1,2,3,4$ HHO formulation in terms of the
kinetic energy relative error $(\mathcal{K}_{0}-\mathcal{K}_{\rm F})/\mathcal{K}_{0}$ and corresponding rates of convergence while halving the diameter of mesh cells. \label{tab:DSL_convercgence_rates}}
\end{table}
In the present numerical investigation, we were able to certify
that the temporal discretization error has a negligible impact on the kinetic energy error, see Table~\ref{tab:DSL_adapt_tolerance}.
Accordingly, we wonder whether the difficulties in reaching the asymptotic convergence rates at the higher polynomial degrees
indicate a lack of regularity of the numerical solution or whether
the achievement of satisfactory asymptotic convergence rates requires finer meshes.
The vorticity field depicted in Figure~\ref{fig:double_shear_layer_vorticity}
shows that extremely high spatial accuracy is required to precisely capture the evolution of the flow field.
Notice, in particular, that the two shear layers evolve into structures that are much smaller than the mesh spacing, even on the finest grid of the mesh sequence.

\begin{figure} [!t]
\centering
\begin{tikzpicture}[scale=0.65]
\begin{axis}    [
                unbounded coords=jump,
                ymode=log,
                scale only axis=true,
                width=1\textwidth,
                height=0.25\textwidth,
                xlabel=time $t$,
                ylabel= time step $\delta t$,
                label style={anchor=near ticklabel, font=\large},
                ticks=major,
                grid=both,
                minor grid style={dashed,gray!30},
                major grid style={solid,gray!50},
                tick pos=left,
                tick align=center,
                xmin=0,
                xmax=2,
                ymin=1E-4,
                ymax=1E-1,
                legend style={cells={anchor=west}},
                legend pos=north west,
                ]
\addplot    [
            thick,
            ]
            table[
            x expr = {\thisrowno{0}},
            y expr = {\thisrowno{1}},
            header=false]
            {DSL_dt_p4_1e-10.dat};
\end{axis}
\end{tikzpicture}
\caption{Double shear layer. Time step size evolution for the $k=4$ HHO formulation over the $\card{\Th}=131072$ grid. \label{fig:double_shear_layer_dt_evolution}}
\end{figure}
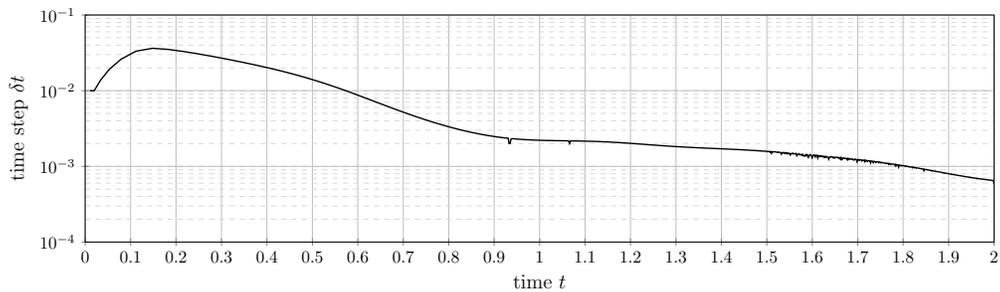

As an insightful indicator of the challenges involved in the double shear layer test case,
the time evolution of the step size $\delta t$ is shown in Figure~\ref{fig:double_shear_layer_dt_evolution}.
Over time, $\delta t$ decreases by nearly two orders of magnitude in order to maintain
the time discretization error within the prescribed tolerance.
Although this variation may seem unremarkable at first glance, notice that,
since we rely on a fifth order accurate time integration strategy, keeping the time step fixed might result in a ten orders of magnitude increase of the temporal discretization error.

\subsection{Ethier--Steinmann} \label{sec:ESs}

In order to numerically assess the spatial convergence rates of the HHO formulation in three space dimensions, we consider the following solution of the Navier--Stokes equations proposed in \cite{EthierSteinman94}:
\begin{align*}
    u(x,y,z,t)  &=  -a \,e^{-d^2t }\,\bigl(e^{ax} \sin(a\,y+z\,d) + e^{za} \cos(a\,x+y\,d)\bigr) , \\
    v(x,y,z,t)  &=  -a \,e^{-d^2t }\,\bigl(e^{ay} \sin(a\,z+x\,d) + e^{xa} \cos(a\,y+z\,d)\bigr) , \\
    w(x,y,z,t)  &=  -a \,e^{-d^2t }\,\bigl(e^{az} \sin(a\,x+y\,d) + e^{ya} \cos(a\,z+x\,d)\bigr) , \\
    r(x,y,z)    &=
                 2     \sin(ax{+}dy) \cos(az{+}dx) e^{a(y{+}z)}
                 {+} 2 \sin(ay{+}dz) \cos(ax{+}dy) e^{a(z{+}x)}
                 {+} 2 \sin(az{+}dx) \cos(ay{+}dz) e^{a(x{+}y)}, \\
    p(x,y,z,t)  &=  -\frac{a^2}{2} \, e^{-2 d^2 t} \, \left(r(x, y, z) + e^{2ax} + e^{2ay} + e^{2az}\right),
\end{align*}
where $u$, $v$ and $w$ are the velocity components in the $x$, $y$ and $z$ directions, respectively, $p$ is the pressure,
$r$ is an auxiliary variable used to define the pressure and $t$ is the time.
The constant parameters are set as $a=\frac{\pi}{4}$ and $d=\frac{\pi}{2}$.
We consider $\nu = \Reynolds^{-1} = 1,0.1,0.01$, as suggested by Ethier and Steinmann in order to avoid wiggles in the proximity of the domain boundary.
The analytic solution is conceived such that the convective term is balanced by the pressure gradient
and the Laplacian of the velocity is balanced by the time derivative.
In order to avoid forcing terms, \ie $\bm{f}=\bm{0}$, the velocity time derivative is multiplied by $\nu$.
The computational domain is the unit cube $\Omega = (0,1)^3$ and time integration is carried out in the time interval $[0,0.1]$.
Accordingly the errors in $L^2$ norm for the velocity, the velocity gradients and the pressure are computed based on the exact solution at time $\tF = 0.1$.
Dirichlet boundary conditions are enforced on five of the six faces composing $\partial \Omega$
while on the top face we impose Neumann boundary conditions.
The velocity and the stress on $\partial \Omega$ are computed relying on the exact solution,
the velocity field is initialized based on the exact solution.

We rely on tetrahedral meshes obtained subdividing in 24 tetrahedrons the hexahedral cells of $N^3$ Cartesian grids,
with $N$ cells per Cartesian direction.
Accordingly, the mesh cardinality is $\card{\Th}=24 N^3$.
The results are reported in Tables \ref{tab:ES_SConvP2}-\ref{tab:ES_SConvP4}
for $k=0,1$ and $k=2,3,4$ HHO formulations, respectively, considering tetrahedral meshes with $N=2i$, $i = 0,\dots,4$.
Notice that, in order to limit the computational burden, the finer meshes are avoided at the higher polynomial degrees.
Time integration is performed relying on a fourth order accurate ESDIRK scheme with time step $\delta t = 0.1/160$.
A $k+1$ rate of error reduction is observed for the pressure error in $L^2$ norm and, interestingly, pressure errors are insentive to a two orders of magnitude reduction of the viscosity.
The velocity and velocity gradient errors in the $L^2$-norm show rates of convergence of $k+2$ and $k+1$, respectively, in the diffusion-dominated regime ($\nu =1$).
Contrary to pressure errors, velocity errors increase while reducing the viscosity, nevertheless the error increase is less pronounced on finer meshes. 
Accordingly, the rate of error reduction for the velocity error overcomes the expected results moving towards the convection-dominated regime.

\begin{table}[!t]
\centering
\small
\begin{tabular}{ccccccccc}
$\nu$         & $\card{\Th}$ & $\| \nabla \uVec_h \|_{L^2(\Omega)}$ & rate &     $\| \uVec_h \|_{L^2(\Omega)}$   & rate   & $\| p_h \|_{L^2(\Omega)}$  & rate   &  $\| \nabla \cdot \uVec_h \|_{L^2(\Omega)}$ \\
\toprule
\multicolumn{9}{c}{$k=0$} \\
\toprule
1          & 24         & 1.0928   & --         & 0.08761  & --    & 0.4572   & --       & 3.57e-16  \\ 
0.1        & 24         & 1.3149   & --         & 0.1381  & --     & 0.3252   & --       & 4.49e-16  \\ 
0.01       & 24         & 4.8507   & --         & 0.7077  & --     & 0.9244   & --       & 1.36e-15  \\ 
\toprule
1          & 192        & 0.5486   & 0.99       & 0.02411   & 1.86  & 0.1852   & 1.30     & 4.06e-16      \\ 
0.1        & 192        & 0.5777   & 1.19       & 0.02886  & 2.26   & 0.1386   & 1.23     & 4.87e-16      \\ 
0.01       & 192        & 1.1356   & 2.09       & 0.08531  & 3.05   & 0.1441   & 2.68     & 4.45e-16      \\ 
\toprule
1          & 1536       & 0.2745   & 1.00       & 0.006139 & 1.97   & 0.07916  & 1.23     & 4.38e-16      \\ 
0.1        & 1536       & 0.2786   & 1.05       & 0.006386 & 2.18   & 0.06709  & 1.05     & 4.16e-16      \\ 
0.01       & 1536       & 0.3932   & 1.53       & 0.01409  & 2.60   & 0.06759  & 1.09     & 4.23e-16     \\ 
\toprule
1          & 12288      & 0.1373   & 1.00       & 0.001547 & 1.99   & 0.03638  & 1.12     & 4.22e-16      \\ 
0.1        & 12288      & 0.1379   & 1.01       & 0.001547 & 2.04   & 0.03332  & 1.01     & 4.17e-16      \\ 
0.01       & 12288      & 0.1640   & 1.26       & 0.002591 & 2.44   & 0.03333  & 1.02     & 4.18e-16      \\ 
\toprule
1          & 98304      & 0.06865  & 1.00       & 0.0003885 & 1.99   & 0.01755  & 1.05    & 4.26e-16      \\ 
0.1        & 98304      & 0.06876  & 1.00       & 0.0003864 & 2.00   & 0.01663  & 1.00    & 4.26e-16      \\ 
0.01       & 98304      & 0.07387  & 1.15       & 0.0004959 & 2.39   & 0.01662  & 1.00    & 4.27e-16      \\ 
\toprule
\multicolumn{9}{c}{$k=1$} \\
\toprule
1          & 24         & 0.1494   & --         & 0.008267 & --         & 0.05984  & --         & 3.93e-16   \\ 
0.1        & 24         & 0.1867    & --         & 0.01143  & --        & 0.05607  & --         & 3.79e-16   \\ 
0.01       & 24         & 0.3793   & --         & 0.02487  & --         & 0.05754  & --         & 4.71e-16   \\ 
\toprule
1          & 192        & 0.03764   & 1.99       & 0.001081  & 2.93     & 0.01802  & 1.73       & 4.38e-16   \\ 
0.1        & 192        & 0.04147  & 2.17       & 0.001249 & 3.19       & 0.01721  & 1.70       & 3.83e-16   \\ 
0.01       & 192        & 0.10791   & 1.81       & 0.004102 & 2.60      & 0.01725  & 1.74       & 3.62e-16   \\ 
\toprule
1          & 1536       & 0.009398 & 2.00       & 0.0001374 & 2.98      & 0.004461  & 2.01      & 4.44e-16   \\ 
0.1        & 1536       & 0.009678 & 2.10       & 0.0001440 & 3.12      & 0.004325 & 1.99       & 4.56e-16   \\ 
0.01       & 1536       & 0.021481  & 2.33       & 0.000427 & 3.26      & 0.004326 & 2.00       & 4.45e-16   \\ 
\toprule
1          & 12288      & 0.002345 & 2.00       & 1.7302e-05 & 2.99     & 0.001108  & 2.01      & 4.29e-16   \\ 
0.1        & 12288      & 0.002356 & 2.04       & 1.7413e-05 & 3.05     & 0.001081 & 2.00       & 4.28e-16   \\ 
0.01       & 12288      & 0.003925  & 2.45       & 3.7014e-05 & 3.53    & 0.001080 & 2.00       & 4.25e-16   \\ 
\toprule
1          & 98304      & 0.0005856 & 2.00       & 2.1708e-06 & 2.99    & 0.0002763 & 2.00      & 4.36e-16   \\ 
0.1        & 98304      & 0.0005849 & 2.01       & 2.1642e-06 & 3.01    & 0.0002702 & 2.00      & 4.38e-16   \\ 
0.01       & 98304      & 0.0007373 & 2.41       & 3.1371e-06 & 3.56    & 0.0002701 & 2.00      & 4.35e-16   \\ 
\bottomrule
\end{tabular}
\caption{Numerical validation of spatial convergence of $k=0$ and $k=1$ HHO formulations, 3D Ethier-Steinman analytical solution of the INS equations.
$L^2(\Omega)$ error norms with respect to the exact solution and the corresponding rates of convergence while halving the mesh step size are reported.
The temporal discretization is chosen such that the temporal approximation error is negligible, see text for details. \label{tab:ES_SConvP2}}
\end{table}

\begin{table}[!t]
\centering
\small
\begin{tabular}{ccccccccc}
$\nu$         & $\card{\Th}$ & $\| \nabla \uVec_h \|_{L^2(\Omega)}$ & rate &     $\| \uVec_h \|_{L^2(\Omega)}$   & rate   & $\| p_h \|_{L^2(\Omega)}$  & rate   &  $\| \nabla \cdot \uVec_h \|_{L^2(\Omega)}$ \\
\toprule
\multicolumn{9}{c}{$k=2$} \\
\toprule
1          & 24         & 0.01856  & --         & 0.0006338 & --           & 0.01507  & --         & 3.65e-16    \\ 
0.1        & 24         & 0.03143  & --         & 0.0012750 & --           & 0.01476  & --         & 4.12e-16    \\ 
0.01       & 24         & 0.08269  & --         & 0.0036416 & --           & 0.01479  & --         & 5.16e-16    \\ 
\toprule
1          & 192        & 0.002247 & 3.05       & 3.9241e-05 & 4.01        & 0.0014929 & 3.34      & 4.14e-16    \\ 
0.1        & 192        & 0.002775 & 3.50       & 5.6648e-05 & 4.49        & 0.0014736 & 3.32      & 4.26e-16    \\ 
0.01       & 192        & 0.007355 & 3.49       & 0.0001868 & 4.28         & 0.0014762  & 3.32     & 4.13e-16    \\ 
\toprule
1          & 1536       & 0.0002809 & 3.00       & 2.4481e-06 & 4.00       & 0.0001771 & 3.08      & 4.54e-16    \\ 
0.1        & 1536       & 0.0002993 & 3.21       & 2.8214e-06 & 4.33       & 0.0001754 & 3.07      & 4.53e-16    \\ 
0.01       & 1536       & 0.0007052 & 3.38       & 8.752e-06  & 4.42       & 0.0001754 & 3.07      & 4.57e-16    \\ 
\toprule
1          & 12288      & 3.5228e-05 & 3.00       & 1.5318e-07 & 4.00      & 2.1918e-05 & 3.01     & 4.52e-16    \\ 
0.1        & 12288      & 3.5534e-05 & 3.07       & 1.5868e-07 & 4.15      & 2.1670e-05 & 3.02     & 4.47e-16    \\ 
0.01       & 12288      & 6.5719e-05 & 3.42       & 3.8828e-07 & 4.49      & 2.1669e-05 & 3.02     & 4.54e-16    \\ 
\toprule
\multicolumn{9}{c}{$k=3$} \\
\toprule
1          & 24         & 0.001642 & --         & 4.3051e-05 & --            & 0.001530 & --         & 3.21e-16  \\ 
0.1        & 24         & 0.002442 & --         & 7.8744e-05 & --            & 0.001514 & --         & 4.86e-16  \\ 
0.01       & 24         & 0.005878 & --         & 0.0001992 & --             & 0.001516 & --         & 3.97e-16  \\ 
\toprule
1          & 192        & 0.0001056 & 3.96       & 1.3451e-06 & 5.00         & 9.9269e-05 & 3.95     & 4.81e-16  \\ 
0.1        & 192        & 0.0001363 & 4.16       & 2.0307e-06 & 5.28         & 9.8632e-05 & 3.94     & 4.53e-16  \\ 
0.01       & 192        & 0.0004248 & 3.79       & 7.0711e-06 & 4.82         & 9.8675e-05 & 3.94     & 4.68e-16  \\ 
\toprule
1          & 1536       & 6.6809e-06 & 3.98       & 4.2303e-08 & 4.99         & 6.2355e-06 & 3.99       & 4.67e-16     \\ 
0.1        & 1536       & 7.1725e-06 & 4.25       & 4.9170e-08 & 5.37         & 6.2005e-06 & 3.99       & 4.73e-16     \\ 
0.01       & 1536       & 2.0616e-05 & 4.37       & 1.7469e-07 & 5.34         & 6.2009e-06 & 3.99       & 4.68e-16     \\ 
\toprule
1          & 12288      & 4.201e-07  & 3.99       & 1.3291e-09 & 4.99         & 4.0745e-07 & 3.94       & 4.72e-16     \\ 
0.1        & 12288      & 4.2330e-07 & 4.08       & 1.3726e-09 & 5.16         & 3.8814e-07 & 4.00       & 4.65e-16     \\ 
0.01       & 12288      & 9.0727e-07 & 4.51       & 3.7794e-09 & 5.53         & 3.8795e-07 & 4.00       & 4.65e-16     \\ 
\toprule
\multicolumn{9}{c}{$k=4$} \\
\toprule
1          & 24         & 0.0001297 & --         & 2.5037e-06 & --         & 0.0001373 & --         & 6.14e-16    \\ 
0.1        & 24         & 0.0004299 & --         & 1.0006e-05 & --         & 0.0001367 & --         & 6.12e-16    \\ 
0.01       & 24         & 0.001314 & --          & 3.1204e-05 & --         & 0.0001371 & --         & 6.30e-16    \\ 
\toprule
1          & 192        & 3.9173e-06 & 5.05      & 3.7850e-08 & 6.05       & 4.4206e-06 & 4.96      & 6.65e-16    \\ 
0.1        & 192        & 6.5719e-06 & 6.03      & 7.8080e-08 & 7.00       & 4.4035e-06 & 4.96      & 6.34e-16    \\ 
0.01       & 192        & 2.6949e-05 & 5.61      & 3.4673e-07 & 6.49       & 4.4063e-06 & 4.96      & 6.79e-16    \\ 
\toprule
1          & 1536       & 1.2390e-07 & 4.98      & 5.9681e-10 & 5.99       & 1.7935e-07 & 4.62      & 6.80e-16    \\ 
0.1        & 1536       & 1.4828e-07 & 5.47      & 8.1797e-10 & 6.58       & 1.3536e-07 & 5.02      & 6.75e-16    \\ 
0.01       & 1536       & 5.8401e-07 & 5.53      & 3.7971e-09 & 6.51       & 1.3486e-07 & 5.03      & 6.76e-16    \\ 
\bottomrule
\end{tabular}
\caption{Numerical validation of spatial convergence of $k=2,3,4$ HHO formulations, 3D Ethier-Steinman analytical solution of the INS equations.
$L^2(\Omega)$ error norms with respect to the exact solution and the corresponding rates of convergence while halving the mesh step size are reported.
The temporal discretization is chosen such that the temporal approximation error is negligible, see text for details. \label{tab:ES_SConvP4}}
\end{table}

\subsection{LLMS pressure gradient test case} \label{sec:LLMS}

The LLMS pressure gradient problem is a manufactured solution for the steady incompressible Stokes equations designed by Lederer, Linke, Merdon, and Schöberl (LLMS) \cite{Lederer.Linke.ea:17} to investigate pressure-robustness.
The HHO formulation has been adapted to this test case dropping the convective term and time derivative.
The analytical velocity and pressure fields read, respectively,
\begin{equation*}
	\uVec	=	\nabla\times \left\lbrace \zeta,\zeta,\zeta \right\rbrace
    \qquad
    \text{and}
    \qquad
    p	=	p_{0}  +x^{5}  +y^{5}  +z^{5},
\end{equation*}
with arbitrary $p_{0}\in\Real$ (here set $p_{0}=1/2$) and $\zeta(x,y,z) = x^{2}\left(x-1\right)^{2}y^{2}\left(y-1\right)^{2}z^{2}\left(z-1\right)^{2}$.
The forcing term $\boldsymbol{f}$ within the momentum equation is set according to the analytical solution.
Dirichlet boundary conditions are imposed on all but one face of the unit cubic domain $\Omega = \left(0,1\right)^3$, 
while a Neumann boundary is enforced on the remaining face.

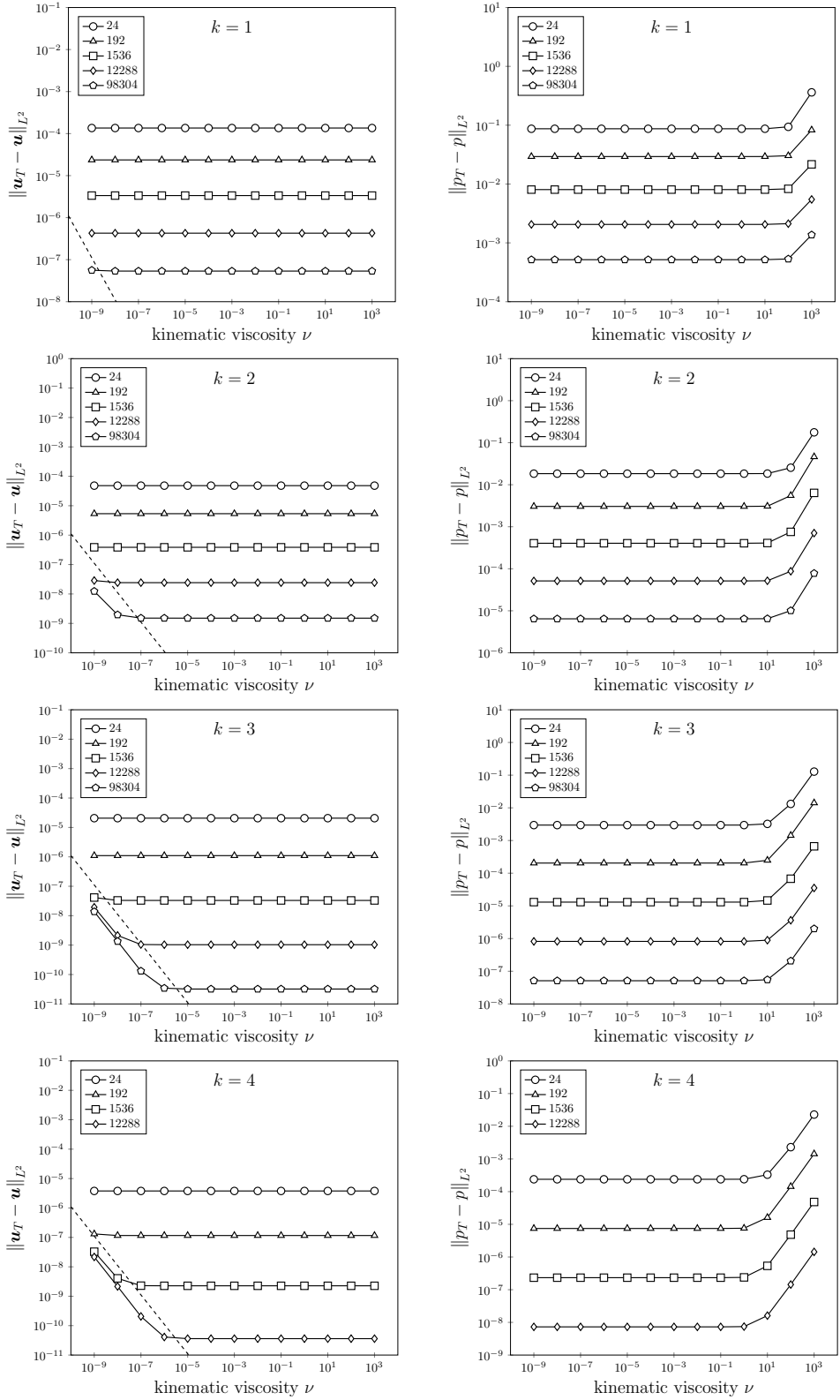
\begin{figure} [!t]
\centering
\begin{tikzpicture}[scale=0.53]
\begin{axis}    [
                unbounded coords=jump,
                ymode=log,xmode=log,
                scale only axis=true,
                width=0.5\textwidth,
                height=0.45\textwidth,
                xlabel= kinematic viscosity $\nu$,
                ylabel= $\left\Vert \uVec_{T}  -\uVec \right\Vert_{L^2}$,
                label style={anchor=near ticklabel, font=\Large},
                ticks=major,
                tick pos=left,
                tick align=center,
                xmin=1.E-10,
                xmax=1.E+4,
                ymin=1E-8,
                ymax=1E-1,
                legend style={cells={anchor=west}},
                legend pos=north west,
                ]
\addplot    [
            thick,
            mark = *,
            mark size=3pt,
            mark options={black, fill=white}
            ]
            table[
            x expr = {\thisrowno{0}},
            y expr = {\thisrowno{2}},
            header=false]
            {LLMS_tetra_p1_24.dat};
\addplot    [
            thick,
            mark = triangle*,
            mark size=3pt,
            mark options={black, fill=white}
            ]
            table[
            x expr = {\thisrowno{0}},
            y expr = {\thisrowno{2}},
            header=false]
            {LLMS_tetra_p1_192.dat};
\addplot    [
            thick,
            mark = square*,
            mark size=3pt,
            mark options={black, fill=white}
            ]
            table[
            x expr = {\thisrowno{0}},
            y expr = {\thisrowno{2}},
            header=false]
            {LLMS_tetra_p1_1536.dat};
\addplot    [
            thick,
            mark = diamond*,
            mark size=3pt,
            mark options={black, fill=white}
            ]
            table[
            x expr = {\thisrowno{0}},
            y expr = {\thisrowno{2}},
            header=false]
            {LLMS_tetra_p1_12288.dat};
\addplot    [
            thick,
            mark = pentagon*,
            mark size=3pt,
            mark options={black, fill=white}
            ]
            table[
            x expr = {\thisrowno{0}},
            y expr = {\thisrowno{2}},
            header=false]
            {LLMS_tetra_p1_98304.dat};
\addplot    [
            thick,
            dashed,
            black,
            domain=1E-10:1E+4] {1.11E-16/x};
\node[anchor=south, font=\Large] at (rel axis cs:0.5,0.9) {$k=1$};
\legend {$24$, $192$, $1536$, $12288$, $98304$}
\end{axis}
\end{tikzpicture}
\hspace{5mm}
\begin{tikzpicture}[scale=0.53]
\begin{axis}    [
                unbounded coords=jump,
                ymode=log,xmode=log,
                scale only axis=true,
                width=0.5\textwidth,
                height=0.45\textwidth,
                xlabel= kinematic viscosity $\nu$,
                ylabel= $\left\Vert p_{T}  -p \right\Vert_{L^2}$,
                label style={anchor=near ticklabel, font=\Large},
                ticks=major,
                tick pos=left,
                tick align=center,
                xmin=1.E-10,
                xmax=1.E+4,
                ymin=1E-4,
                ymax=1E+1,
                legend style={cells={anchor=west}},
                legend pos=north west,
                ]
\addplot    [
            thick,
            mark = *,
            mark size=3pt,
            mark options={black, fill=white}
            ]
            table[
            x expr = {\thisrowno{0}},
            y expr = {\thisrowno{3}},
            header=false]
            {LLMS_tetra_p1_24.dat};
\addplot    [
            thick,
            mark = triangle*,
            mark size=3pt,
            mark options={black, fill=white}
            ]
            table[
            x expr = {\thisrowno{0}},
            y expr = {\thisrowno{3}},
            header=false]
            {LLMS_tetra_p1_192.dat};
\addplot    [
            thick,
            mark = square*,
            mark size=3pt,
            mark options={black, fill=white}
            ]
            table[
            x expr = {\thisrowno{0}},
            y expr = {\thisrowno{3}},
            header=false]
            {LLMS_tetra_p1_1536.dat};
\addplot    [
            thick,
            mark = diamond*,
            mark size=3pt,
            mark options={black, fill=white}
            ]
            table[
            x expr = {\thisrowno{0}},
            y expr = {\thisrowno{3}},
            header=false]
            {LLMS_tetra_p1_12288.dat};
\addplot    [
            thick,
            mark = pentagon*,
            mark size=3pt,
            mark options={black, fill=white}
            ]
            table[
            x expr = {\thisrowno{0}},
            y expr = {\thisrowno{3}},
            header=false]
            {LLMS_tetra_p1_98304.dat};
\node[anchor=south, font=\Large] at (rel axis cs:0.5,0.9) {$k=1$};
\legend {$24$, $192$, $1536$, $12288$, $98304$}
\end{axis}
\end{tikzpicture}
\\
\begin{tikzpicture}[scale=0.53]
\begin{axis}    [
                unbounded coords=jump,
                ymode=log,xmode=log,
                scale only axis=true,
                width=0.5\textwidth,
                height=0.45\textwidth,
                xlabel= kinematic viscosity $\nu$,
                ylabel= $\left\Vert \uVec_{T}  -\uVec \right\Vert_{L^2}$,
                label style={anchor=near ticklabel, font=\Large},
                ticks=major,
                tick pos=left,
                tick align=center,
                xmin=1.E-10,
                xmax=1.E+4,
                ymin=1E-10,
                ymax=1E-0,
                legend style={cells={anchor=west}},
                legend pos=north west,
                ]
\addplot    [
            thick,
            mark = *,
            mark size=3pt,
            mark options={black, fill=white}
            ]
            table[
            x expr = {\thisrowno{0}},
            y expr = {\thisrowno{2}},
            header=false]
            {LLMS_tetra_p2_24.dat};
\addplot    [
            thick,
            mark = triangle*,
            mark size=3pt,
            mark options={black, fill=white}
            ]
            table[
            x expr = {\thisrowno{0}},
            y expr = {\thisrowno{2}},
            header=false]
            {LLMS_tetra_p2_192.dat};
\addplot    [
            thick,
            mark = square*,
            mark size=3pt,
            mark options={black, fill=white}
            ]
            table[
            x expr = {\thisrowno{0}},
            y expr = {\thisrowno{2}},
            header=false]
            {LLMS_tetra_p2_1536.dat};
\addplot    [
            thick,
            mark = diamond*,
            mark size=3pt,
            mark options={black, fill=white}
            ]
            table[
            x expr = {\thisrowno{0}},
            y expr = {\thisrowno{2}},
            header=false]
            {LLMS_tetra_p2_12288.dat};
\addplot    [
            thick,
            mark = pentagon*,
            mark size=3pt,
            mark options={black, fill=white}
            ]
            table[
            x expr = {\thisrowno{0}},
            y expr = {\thisrowno{2}},
            header=false]
            {LLMS_tetra_p2_98304.dat};
\addplot    [
            thick,
            dashed,
            black,
            domain=1E-10:1E+4] {1.11E-16/x};
\node[anchor=south, font=\Large] at (rel axis cs:0.5,0.9) {$k=2$};
\legend {$24$, $192$, $1536$, $12288$, $98304$}
\end{axis}
\end{tikzpicture}
\hspace{5mm}
\begin{tikzpicture}[scale=0.53]
\begin{axis}    [
                unbounded coords=jump,
                ymode=log,xmode=log,
                scale only axis=true,
                width=0.5\textwidth,
                height=0.45\textwidth,
                xlabel= kinematic viscosity $\nu$,
                ylabel= $\left\Vert p_{T}  -p \right\Vert_{L^2}$,
                label style={anchor=near ticklabel, font=\Large},
                ticks=major,
                tick pos=left,
                tick align=center,
                xmin=1.E-10,
                xmax=1.E+4,
                ymin=1E-6,
                ymax=1E+1,
                legend style={cells={anchor=west}},
                legend pos=north west,
                ]
\addplot    [
            thick,
            mark = *,
            mark size=3pt,
            mark options={black, fill=white}
            ]
            table[
            x expr = {\thisrowno{0}},
            y expr = {\thisrowno{3}},
            header=false]
            {LLMS_tetra_p2_24.dat};
\addplot    [
            thick,
            mark = triangle*,
            mark size=3pt,
            mark options={black, fill=white}
            ]
            table[
            x expr = {\thisrowno{0}},
            y expr = {\thisrowno{3}},
            header=false]
            {LLMS_tetra_p2_192.dat};
\addplot    [
            thick,
            mark = square*,
            mark size=3pt,
            mark options={black, fill=white}
            ]
            table[
            x expr = {\thisrowno{0}},
            y expr = {\thisrowno{3}},
            header=false]
            {LLMS_tetra_p2_1536.dat};
\addplot    [
            thick,
            mark = diamond*,
            mark size=3pt,
            mark options={black, fill=white}
            ]
            table[
            x expr = {\thisrowno{0}},
            y expr = {\thisrowno{3}},
            header=false]
            {LLMS_tetra_p2_12288.dat};
\addplot    [
            thick,
            mark = pentagon*,
            mark size=3pt,
            mark options={black, fill=white}
            ]
            table[
            x expr = {\thisrowno{0}},
            y expr = {\thisrowno{3}},
            header=false]
            {LLMS_tetra_p2_98304.dat};
\node[anchor=south, font=\Large] at (rel axis cs:0.5,0.9) {$k=2$};
\legend {$24$, $192$, $1536$, $12288$, $98304$}
\end{axis}
\end{tikzpicture}
\\
\begin{tikzpicture}[scale=0.53]
\begin{axis}    [
                unbounded coords=jump,
                ymode=log,xmode=log,
                scale only axis=true,
                width=0.5\textwidth,
                height=0.45\textwidth,
                xlabel= kinematic viscosity $\nu$,
                ylabel= $\left\Vert \uVec_{T}  -\uVec \right\Vert_{L^2}$,
                label style={anchor=near ticklabel, font=\Large},
                ticks=major,
                tick pos=left,
                tick align=center,
                xmin=1.E-10,
                xmax=1.E+4,
                ymin=1E-11,
                ymax=1E-1,
                legend style={cells={anchor=west}},
                legend pos=north west,
                ]
\addplot    [
            thick,
            mark = *,
            mark size=3pt,
            mark options={black, fill=white}
            ]
            table[
            x expr = {\thisrowno{0}},
            y expr = {\thisrowno{2}},
            header=false]
            {LLMS_tetra_p3_24.dat};
\addplot    [
            thick,
            mark = triangle*,
            mark size=3pt,
            mark options={black, fill=white}
            ]
            table[
            x expr = {\thisrowno{0}},
            y expr = {\thisrowno{2}},
            header=false]
            {LLMS_tetra_p3_192.dat};
\addplot    [
            thick,
            mark = square*,
            mark size=3pt,
            mark options={black, fill=white}
            ]
            table[
            x expr = {\thisrowno{0}},
            y expr = {\thisrowno{2}},
            header=false]
            {LLMS_tetra_p3_1536.dat};
\addplot    [
            thick,
            mark = diamond*,
            mark size=3pt,
            mark options={black, fill=white}
            ]
            table[
            x expr = {\thisrowno{0}},
            y expr = {\thisrowno{2}},
            header=false]
            {LLMS_tetra_p3_12288.dat};
\addplot    [
            thick,
            mark = pentagon*,
            mark size=3pt,
            mark options={black, fill=white}
            ]
            table[
            x expr = {\thisrowno{0}},
            y expr = {\thisrowno{2}},
            header=false]
            {LLMS_tetra_p3_98304.dat};
\addplot    [
            thick,
            dashed,
            black,
            domain=1E-10:1E+4] {1.11E-16/x};

\node[anchor=south, font=\Large] at (rel axis cs:0.5,0.9) {$k=3$};
\legend {$24$, $192$, $1536$, $12288$, $98304$}
\end{axis}
\end{tikzpicture}
\hspace{5mm}
\begin{tikzpicture}[scale=0.53]
\begin{axis}    [
                unbounded coords=jump,
                ymode=log,xmode=log,
                scale only axis=true,
                width=0.5\textwidth,
                height=0.45\textwidth,
                xlabel= kinematic viscosity $\nu$,
                ylabel= $\left\Vert p_{T}  -p \right\Vert_{L^2}$,
                label style={anchor=near ticklabel, font=\Large},
                ticks=major,
                tick pos=left,
                tick align=center,
                xmin=1.E-10,
                xmax=1.E+4,
                ymin=1E-8,
                ymax=1E+1,
                legend style={cells={anchor=west}},
                legend pos=north west,
                ]
\addplot    [
            thick,
            mark = *,
            mark size=3pt,
            mark options={black, fill=white}
            ]
            table[
            x expr = {\thisrowno{0}},
            y expr = {\thisrowno{3}},
            header=false]
            {LLMS_tetra_p3_24.dat};
\addplot    [
            thick,
            mark = triangle*,
            mark size=3pt,
            mark options={black, fill=white}
            ]
            table[
            x expr = {\thisrowno{0}},
            y expr = {\thisrowno{3}},
            header=false]
            {LLMS_tetra_p3_192.dat};
\addplot    [
            thick,
            mark = square*,
            mark size=3pt,
            mark options={black, fill=white}
            ]
            table[
            x expr = {\thisrowno{0}},
            y expr = {\thisrowno{3}},
            header=false]
            {LLMS_tetra_p3_1536.dat};
\addplot    [
            thick,
            mark = diamond*,
            mark size=3pt,
            mark options={black, fill=white}
            ]
            table[
            x expr = {\thisrowno{0}},
            y expr = {\thisrowno{3}},
            header=false]
            {LLMS_tetra_p3_12288.dat};
\addplot    [
            thick,
            mark = pentagon*,
            mark size=3pt,
            mark options={black, fill=white}
            ]
            table[
            x expr = {\thisrowno{0}},
            y expr = {\thisrowno{3}},
            header=false]
            {LLMS_tetra_p3_98304.dat};
\node[anchor=south, font=\Large] at (rel axis cs:0.5,0.9) {$k=3$};
\legend {$24$, $192$, $1536$, $12288$, $98304$}
\end{axis}
\end{tikzpicture}
\\
\begin{tikzpicture}[scale=0.53]
\begin{axis}    [
                unbounded coords=jump,
                ymode=log,xmode=log,
                scale only axis=true,
                width=0.5\textwidth,
                height=0.45\textwidth,
                xlabel= kinematic viscosity $\nu$,
                ylabel= $\left\Vert \uVec_{T}  -\uVec \right\Vert_{L^2}$,
                label style={anchor=near ticklabel, font=\Large},
                ticks=major,
                tick pos=left,
                tick align=center,
                xmin=1.E-10,
                xmax=1.E+4,
                ymin=1E-11,
                ymax=1E-1,
                legend style={cells={anchor=west}},
                legend pos=north west,
                ]
\addplot    [
            thick,
            mark = *,
            mark size=3pt,
            mark options={black, fill=white}
            ]
            table[
            x expr = {\thisrowno{0}},
            y expr = {\thisrowno{2}},
            header=false]
            {LLMS_tetra_p4_24.dat};
\addplot    [
            thick,
            mark = triangle*,
            mark size=3pt,
            mark options={black, fill=white}
            ]
            table[
            x expr = {\thisrowno{0}},
            y expr = {\thisrowno{2}},
            header=false]
            {LLMS_tetra_p4_192.dat};
\addplot    [
            thick,
            mark = square*,
            mark size=3pt,
            mark options={black, fill=white}
            ]
            table[
            x expr = {\thisrowno{0}},
            y expr = {\thisrowno{2}},
            header=false]
            {LLMS_tetra_p4_1536.dat};
\addplot    [
            thick,
            mark = diamond*,
            mark size=3pt,
            mark options={black, fill=white}
            ]
            table[
            x expr = {\thisrowno{0}},
            y expr = {\thisrowno{2}},
            header=false]
            {LLMS_tetra_p4_12288.dat};
\addplot    [
            thick,
            dashed,
            black,
            domain=1E-10:1E+4] {1.11E-16/x};
\node[anchor=south, font=\Large] at (rel axis cs:0.5,0.9) {$k=4$};
\legend {$24$, $192$, $1536$, $12288$}
\end{axis}
\end{tikzpicture}
\hspace{5mm}
\begin{tikzpicture}[scale=0.53]
\begin{axis}    [
                unbounded coords=jump,
                ymode=log,xmode=log,
                scale only axis=true,
                width=0.5\textwidth,
                height=0.45\textwidth,
                xlabel= kinematic viscosity $\nu$,
                ylabel= $\left\Vert p_{T}  -p \right\Vert_{L^2}$,
                label style={anchor=near ticklabel, font=\Large},
                ticks=major,
                tick pos=left,
                tick align=center,
                xmin=1.E-10,
                xmax=1.E+4,
                ymin=1E-9,
                ymax=1E+0,
                legend style={cells={anchor=west}},
                legend pos=north west,
                ]
\addplot    [
            thick,
            mark = *,
            mark size=3pt,
            mark options={black, fill=white}
            ]
            table[
            x expr = {\thisrowno{0}},
            y expr = {\thisrowno{3}},
            header=false]
            {LLMS_tetra_p4_24.dat};
\addplot    [
            thick,
            mark = triangle*,
            mark size=3pt,
            mark options={black, fill=white}
            ]
            table[
            x expr = {\thisrowno{0}},
            y expr = {\thisrowno{3}},
            header=false]
            {LLMS_tetra_p4_192.dat};
\addplot    [
            thick,
            mark = square*,
            mark size=3pt,
            mark options={black, fill=white}
            ]
            table[
            x expr = {\thisrowno{0}},
            y expr = {\thisrowno{3}},
            header=false]
            {LLMS_tetra_p4_1536.dat};
\addplot    [
            thick,
            mark = diamond*,
            mark size=3pt,
            mark options={black, fill=white}
            ]
            table[
            x expr = {\thisrowno{0}},
            y expr = {\thisrowno{3}},
            header=false]
            {LLMS_tetra_p4_12288.dat};
\node[anchor=south, font=\Large] at (rel axis cs:0.5,0.9) {$k=4$};
\legend {$24$, $192$, $1536$, $12288$}
\end{axis}
\end{tikzpicture}
\caption{LLMS Stokes problem test case. \textit{Left and right}: velocity and pressure errors varying the viscosity.
We consider $k=1,2,3,4$ HHO formulations. The figure legend shows the grid cardinalities.
The black dashed line depicts $\epsilon/\nu$, where $\epsilon = 1.11e-16$ is the double-precision floating-point machine epsilon. \label{fig:LLMS_tets}}
\end{figure}

The pressure-robustness capabilities of the HHO formulation are evaluated on an $h$-refined sequence of meshes with cardinalities $\card{\Th} = 24 \times 8^{i}, \, i = 0, \dots, 4$, consisting of regular tetrahedral cells.
Results for $k =1,2,3,4$ HHO formulations are presented in Figure~\ref{fig:LLMS_tets} considering a wide range of viscosities $\nu=10^{j}$, $j=-9,\dots,3$.
Given a polynomial approximation and a mesh cardinality, $\boldsymbol{u}_h$ remains identical for all viscosity values.
Since, while decreasing $\nu$, the pressure error starts to dominate, this is a clear indication that velocity error is insensitive to the pressure error.
This behavior clearly demonstrates the pressure-robust nature of the formulation considered.
We remark that, when considering the smallest viscosity values and the highest spatial discretization accuracies,
the influence of finite arithmetic precision comes into play.
The velocity error grows with a factor $\epsilon/\nu$ (see the dashed line in the bottom left corner of figures), where $\epsilon$ is the double-precision floating-point machine epsilon.
This behavior can be explained by noticing that the Stokes problem is ill-posed in the inviscid limit,
and, accordingly, it is expected that the velocity error grows unbounded approaching $\nu=0$.
In contrast, since the velocity error dominates when the viscosity is big enough, also the pressure error increases for viscosity values bigger than one.

\begin{figure} [!t]
\centering
\begin{tikzpicture}[scale=0.53]
\begin{axis}    [
                unbounded coords=jump,
                ymode=log,xmode=log,
                scale only axis=true,
                width=0.5\textwidth,
                height=0.45\textwidth,
                xlabel= kinematic viscosity $\nu$,
                ylabel= $\left\Vert \uVec_{T}  -\uVec \right\Vert_{L^2}$,
                label style={anchor=near ticklabel, font=\Large},
                ticks=major,
                tick pos=left,
                tick align=center,
                xmin=1.E-10,
                xmax=1.E+4,
                ymin=1E-8,
                ymax=1E-1,
                legend style={cells={anchor=west}},
                legend pos=north west,
                ]
\addplot    [
            thick,
            mark = *,
            mark size=3pt,
            mark options={black, fill=white}
            ]
            table[
            x expr = {\thisrowno{0}},
            y expr = {\thisrowno{2}},
            header=false]
            {LLMS_pyram_p2_6.dat};
\addplot    [
            thick,
            mark = triangle*,
            mark size=3pt,
            mark options={black, fill=white}
            ]
            table[
            x expr = {\thisrowno{0}},
            y expr = {\thisrowno{2}},
            header=false]
            {LLMS_pyram_p2_48.dat};
\addplot    [
            thick,
            mark = square*,
            mark size=3pt,
            mark options={black, fill=white}
            ]
            table[
            x expr = {\thisrowno{0}},
            y expr = {\thisrowno{2}},
            header=false]
            {LLMS_pyram_p2_384.dat};
\addplot    [
            thick,
            mark = diamond*,
            mark size=3pt,
            mark options={black, fill=white}
            ]
            table[
            x expr = {\thisrowno{0}},
            y expr = {\thisrowno{2}},
            header=false]
            {LLMS_pyram_p2_3072.dat};
\addplot    [
            thick,
            mark = pentagon*,
            mark size=3pt,
            mark options={black, fill=white}
            ]
            table[
            x expr = {\thisrowno{0}},
            y expr = {\thisrowno{2}},
            header=false]
            {LLMS_pyram_p2_24576.dat};
\addplot    [
            thick,
            dashed,
            black,
            domain=1E-10:1E+4] {1.11E-16/x};
\node[anchor=south, font=\Large] at (rel axis cs:0.5,0.9) {pyramid, $k=2$};
\legend {$6$, $48$, $384$, $3072$, $24576$}
\end{axis}
\end{tikzpicture}
\hspace{5mm}
\begin{tikzpicture}[scale=0.53]
\begin{axis}    [
                unbounded coords=jump,
                ymode=log,xmode=log,
                scale only axis=true,
                width=0.5\textwidth,
                height=0.45\textwidth,
                xlabel= kinematic viscosity $\nu$,
                ylabel= $\left\Vert p_{T}  -p \right\Vert_{L^2}$,
                label style={anchor=near ticklabel, font=\Large},
                ticks=major,
                tick pos=left,
                tick align=center,
                xmin=1.E-10,
                xmax=1.E+4,
                ymin=1E-5,
                ymax=1E+1,
                legend style={cells={anchor=west}},
                legend pos=north west,
                ]
\addplot    [
            thick,
            mark = *,
            mark size=3pt,
            mark options={black, fill=white}
            ]
            table[
            x expr = {\thisrowno{0}},
            y expr = {\thisrowno{3}},
            header=false]
            {LLMS_pyram_p2_6.dat};
\addplot    [
            thick,
            mark = triangle*,
            mark size=3pt,
            mark options={black, fill=white}
            ]
            table[
            x expr = {\thisrowno{0}},
            y expr = {\thisrowno{3}},
            header=false]
            {LLMS_pyram_p2_48.dat};
\addplot    [
            thick,
            mark = square*,
            mark size=3pt,
            mark options={black, fill=white}
            ]
            table[
            x expr = {\thisrowno{0}},
            y expr = {\thisrowno{3}},
            header=false]
            {LLMS_pyram_p2_384.dat};
\addplot    [
            thick,
            mark = diamond*,
            mark size=3pt,
            mark options={black, fill=white}
            ]
            table[
            x expr = {\thisrowno{0}},
            y expr = {\thisrowno{3}},
            header=false]
            {LLMS_pyram_p2_3072.dat};
\addplot    [
            thick,
            mark = pentagon*,
            mark size=3pt,
            mark options={black, fill=white}
            ]
            table[
            x expr = {\thisrowno{0}},
            y expr = {\thisrowno{3}},
            header=false]
            {LLMS_pyram_p2_24576.dat};
\node[anchor=south, font=\Large] at (rel axis cs:0.5,0.9) {pyramid, $k=2$};
\legend {$6$, $48$, $384$, $3072$, $24576$}
\end{axis}
\end{tikzpicture}
\\
\begin{tikzpicture}[scale=0.53]
\begin{axis}    [
                unbounded coords=jump,
                ymode=log,xmode=log,
                scale only axis=true,
                width=0.5\textwidth,
                height=0.45\textwidth,
                xlabel= kinematic viscosity $\nu$,
                ylabel= $\left\Vert \uVec_{T}  -\uVec \right\Vert_{L^2}$,
                label style={anchor=near ticklabel, font=\Large},
                ticks=major,
                tick pos=left,
                tick align=center,
                xmin=1.E-10,
                xmax=1.E+4,
                ymin=1E-8,
                ymax=1E-1,
                legend style={cells={anchor=west}},
                legend pos=north west,
                ]
\addplot    [
            thick,
            mark = *,
            mark size=3pt,
            mark options={black, fill=white}
            ]
            table[
            x expr = {\thisrowno{0}},
            y expr = {\thisrowno{2}},
            header=false]
            {LLMS_prism_p2_16.dat};
\addplot    [
            thick,
            mark = triangle*,
            mark size=3pt,
            mark options={black, fill=white}
            ]
            table[
            x expr = {\thisrowno{0}},
            y expr = {\thisrowno{2}},
            header=false]
            {LLMS_prism_p2_128.dat};
\addplot    [
            thick,
            mark = square*,
            mark size=3pt,
            mark options={black, fill=white}
            ]
            table[
            x expr = {\thisrowno{0}},
            y expr = {\thisrowno{2}},
            header=false]
            {LLMS_prism_p2_1024.dat};
\addplot    [
            thick,
            mark = diamond*,
            mark size=3pt,
            mark options={black, fill=white}
            ]
            table[
            x expr = {\thisrowno{0}},
            y expr = {\thisrowno{2}},
            header=false]
            {LLMS_prism_p2_8192.dat};
\addplot    [
            thick,
            mark = pentagon*,
            mark size=3pt,
            mark options={black, fill=white}
            ]
            table[
            x expr = {\thisrowno{0}},
            y expr = {\thisrowno{2}},
            header=false]
            {LLMS_prism_p2_65536.dat};
\addplot    [
            thick,
            dashed,
            black,
            domain=1E-10:1E+4] {1.11E-16/x};

\node[anchor=south, font=\Large] at (rel axis cs:0.5,0.9) {prism, $k=2$};
\legend {$16$, $128$, $1024$, $8192$, $65536$}
\end{axis}
\end{tikzpicture}
\hspace{5mm}
\begin{tikzpicture}[scale=0.53]
\begin{axis}    [
                unbounded coords=jump,
                ymode=log,xmode=log,
                scale only axis=true,
                width=0.5\textwidth,
                height=0.45\textwidth,
                xlabel= kinematic viscosity $\nu$,
                ylabel= $\left\Vert p_{T}  -p \right\Vert_{L^2}$,
                label style={anchor=near ticklabel, font=\Large},
                ticks=major,
                tick pos=left,
                tick align=center,
                xmin=1.E-10,
                xmax=1.E+4,
                ymin=1E-6,
                ymax=1E+1,
                legend style={cells={anchor=west}},
                legend pos=north west,
                ]
\addplot    [
            thick,
            mark = *,
            mark size=3pt,
            mark options={black, fill=white}
            ]
            table[
            x expr = {\thisrowno{0}},
            y expr = {\thisrowno{3}},
            header=false]
            {LLMS_prism_p2_16.dat};
\addplot    [
            thick,
            mark = triangle*,
            mark size=3pt,
            mark options={black, fill=white}
            ]
            table[
            x expr = {\thisrowno{0}},
            y expr = {\thisrowno{3}},
            header=false]
            {LLMS_prism_p2_128.dat};
\addplot    [
            thick,
            mark = square*,
            mark size=3pt,
            mark options={black, fill=white}
            ]
            table[
            x expr = {\thisrowno{0}},
            y expr = {\thisrowno{3}},
            header=false]
            {LLMS_prism_p2_1024.dat};
\addplot    [
            thick,
            mark = diamond*,
            mark size=3pt,
            mark options={black, fill=white}
            ]
            table[
            x expr = {\thisrowno{0}},
            y expr = {\thisrowno{3}},
            header=false]
            {LLMS_prism_p2_8192.dat};
\addplot    [
            thick,
            mark = pentagon*,
            mark size=3pt,
            mark options={black, fill=white}
            ]
            table[
            x expr = {\thisrowno{0}},
            y expr = {\thisrowno{3}},
            header=false]
            {LLMS_prism_p2_65536.dat};
\node[anchor=south, font=\Large] at (rel axis cs:0.5,0.9) {prism, $k=2$};
\legend {$16$, $128$, $1024$, $8192$, $65536$}
\end{axis}
\end{tikzpicture}
\caption{LLMS Stokes problem test case.
\textit{Left and right}: velocity and pressure errors in $L^2$ norm varying the viscosity.
$k=2$ HHO formulations on pyramidal (\textit{top}), and prismatic (\textit{bottom}) mesh cells are considered.
The figure legend shows grid cardinalities. The black dashed line depicts $\epsilon/\nu$, where $\epsilon = 1.11 \times 10^{-16}$ is the double-precision floating-point machine epsilon. \label{fig:LLMS_p2_tets_pyrs_pris}}
\end{figure}
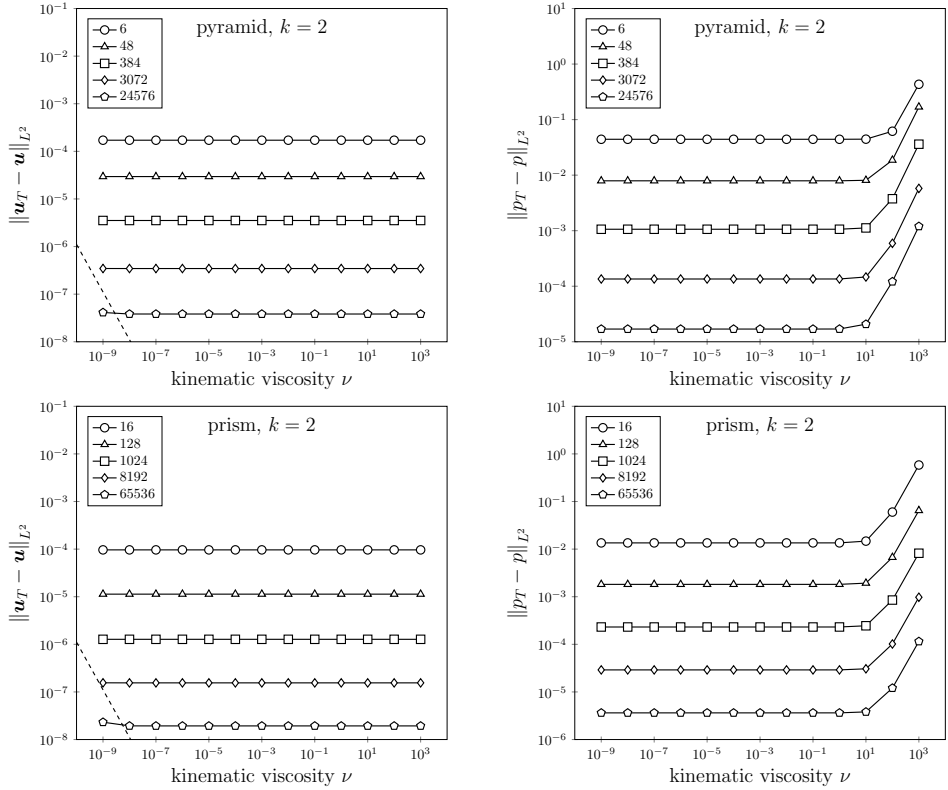

\paragraph{Remark on non-simplicial meshes} \label{sec:remarkNonSimp}
In this work all test cases but the ones presented in the last portion of this section consider simplicial meshes.
Indeed, the theoretical analysis carried out in \cite{BBDiPietroMassa25} shows that, 
when considering non-simplicial meshes, different choices of polynomial spaces should be adopted to maintain optimal convergence rates.
In practice, as shown in Figure~\ref{fig:LLMS_p2_tets_pyrs_pris}, the HHO formulation here proposed can be employed as is on prismatic and
pyramidal cells maintaining pressure robustness, but losing one order of convergence for the velocity and velocity gradients error in $L^2$ norm.
This means that the expected rate of convergences for HHO discretization of the Stokes problem on simplicial and non-simplicial meshes is $k{+}2$ and $k{+}1$, respectively, for the velocity error in $L^2$-norm.
The pressure error in the $L^2$-norm shows a convergence rate of $k{+}1$ on both simplicial and non-simplicial meshes.
Basically, when considering the incompressible Navier--Stokes problem on non-simplicial meshes, one can expect the same convergence rates in the viscous dominated and convection-dominated regimes.

The investigation of pressure robustness reported in Figure~\ref{fig:LLMS_p2_tets_pyrs_pris} considers $k=2$ HHO formulations and
two regular mesh sequences of cardinality $\card{\Th}= 6 \times 8^{i}$, and $\card{\Th}= 16 \times 8^{i}$, $i = 0, \dots, 4$, respectively, in case of pyramidal and prismatic mesh cells.
The numerical results confirm insensitivity of the velocity field to the irrotational part of body forces.


\section{Application to turbulence modelling}\label{sec:turbulence}

The extensive numerical validation conducted in the previous section suggests that the high-order accurate ESDIRK-HHO formulation
should be capable of tackling incompressible flow problems in the turbulent flow regime.
To this end we consider the well known Taylor--Green vortex (TGV) problem at Reynolds $\Reynolds = 1\,600$.
The Taylor--Green vortex problem has been widely adopted to challenge the turbulence modelling capabilities of numerical schemes \cite{HDGHDIVTGV19,FEHN2018667} because,
despite its simple setup, proper handling of the energy cascade phenomena is required in order to precisely replicate the evolution of vortical structures.
Moreover, the solver performance can be assessed comparing with DNS data available in literature.
We use here the results of \cite{VANREES20112794}.

All the TGV computations have been performed with an older version of the code whose implementation omits $L^2$ projections of test functions in the convective term formulation, see paragraph \ref{sec:SpaceDiscr}.
While the convergence rates are slightly suboptimal (half an order is lost), the impact in the context of TGV is marginal at the higher polynomials degrees.
Due to the computationally intensive nature of the simulations, it has not been possible to rerun all tests with the updated code implementation.

The computational domain is the triple periodic cube $\Omega = (-\pi,\pi)^3$
and the initial velocity and pressure fields are defined as follows:
\begin{align*}
    u(x,y,z)  &= \sin(x)\cos(y)\cos(z), \\
    v(x,y,z)  &=-\cos(x)\sin(y)\cos(z) , \\
    w(x,y,z)  &= 0, \\
    p(x,y,z)  &= 1+\bigl(\cos(2x)+\cos(2y)\bigr) \, \bigl(\cos(2z)+2\bigr)/16.
\end{align*}
where $u$, $v$ and $w$ are the velocity components in the $x$, $y$ and $z$ directions, respectively,
and $p$ is the pressure.
Time integration is performed from $t_0 = 0$ to $\tF = 20$ with $\bm{f} = \bm{0}$.

Since the cubic domain is periodic, all the terms written in conservative form
(as the divergence of a vector field) can be eliminated from physical laws expressing conservation principles.
Accordingly, the kinetic energy equation
\[
\frac{d}{dt} \int_\Omega \frac12 \uVec \cdot \uVec
= -\nu \int_\Omega (\nabla \times \uVec) \cdot (\nabla \times \uVec),
\]
states that time evolution of the kinetic energy is driven
by the enstrophy scaled by $\nu={1}/{\Reynolds}$, which is the only form of energy dissipation.
Accordingly, at each point in time, the most effective way of quantifying the accuracy of an approximated velocity field consist in
\begin{inparaenum}[i)]
\item comparing the average enstrophy over $\Omega$ with DNS, and \item assessing the
discrepancy, if any, between the time derivative of the average kinetic energy and the average dissipation.
\end{inparaenum}
Clearly, the second analysis checks weather the balance of kinetic energy is properly replicated at the discrete level.
Let $\uVec_h \in L^2(\Omega)^d$ be the discrete velocity solution at time $t$, obtained by gluing together the cell components.
Introducing the physical dissipation 
$\mathcal{E} = \nu \int_\Omega (\nabla \times \uVec_h) \cdot (\nabla \times \uVec_h)$
and the total dissipation
$\mathcal{E}_h = - \frac{d}{dt} \int_\Omega \frac{1}{2} (\uVec_h \cdot \uVec_h)$,
the numerical dissipation introduced by the spatial-temporal discretization reads $\mathcal{E}_h - \mathcal{E}$.

We solve the problem on two tetrahedral elements grids: $\card{\Th} = 24 \times (4i)^3$, with $i=1,2$.
The first ($i{=}1$) and the second ($i{=}2$) grid are generated starting 
from Cartesian meshes consisting of $4$ and $8$ hexahedral cells per direction, respectively, 
and subdividing each hexahedral cell into 24 tetrahedrons.
On both those meshes, we consider ESDIRK-HHO formulations with $k=1,\dots,9$ and fifth order accurate time integration with local time step adaptation.
On the finer grid $k=9$, computations are missing because of lack of adequate computational resources.

In Table~\ref{tab:TGV_dofs} the total number of degrees of freedom (DOF) are reported for each combination of computational grid and polynomial degree $k$.
We distinguish among cells and faces degrees of freedom, computed as follows 
$$
{\rm DOF\textsubscript{T}}=\card{\Th}\left(3\dim(\Polyd{3}{k+1})+\dim(\Polyd{3}{k})\right), \quad \rm{and} \quad
{\rm DOF\textsubscript{F}}=\card{\Mh}\left(3\dim(\Polyd{2}{k})+\dim(\Polyd{2}{k+1})\right),
$$
respectively.
Notice that, thanks to static condensation, the global matrix dimension is DOF\textsubscript{F}.
Additionally, since Cartesian element meshes are commonly employed for this test case, 
we provide, for the sake of comparison with reference DNS data, the equivalent number of degrees of 
freedom per direction for each velocity component. Once again, we distinguish 
among cells and faces equivalent degrees of freedom, computed as follows:
$$
{\rm eDOFu{-}1D\textsubscript{T}}=\left(\card{\Th}\dim(\Polyd{3}{k+1})\right)^{\frac{1}{3}}, \quad \rm{and} \quad
{\rm eDOFu{-}1D\textsubscript{F}}=\left(\card{\Mh}\dim(\Polyd{2}{k})\right)^{\frac{1}{3}}, 
$$
respectively.
DNS data is obtained based on a spatial resultion of $512$ degrees of freedom per Cartesian direction for each velocity component.
\begin{table}[!t]
\centering
\small
\begin{tabular}{cc}
\begin{tabular}[t]{c cccc}
& \multicolumn{4}{c}{grid $24 \times 4^3$} \\
\toprule
$k$ & DOF\textsubscript{T} & DOF\textsubscript{F} & eDOFu-1D\textsubscript{T} & eDOFu-1D\textsubscript{F} \\
\midrule
1 &  15360 &  9216 & 25 & 21 \\ 
2 &  30720 & 18432 & 31 & 26 \\ 
3 &  53760 & 30720 & 38 & 31 \\ 
4 &  86016 & 46080 & 44 & 36 \\ 
5 & 129024 & 64512 & 51 & 40 \\ 
6 & 184320 & 86016 & 57 & 44 \\ 
7 & 253440 &110592 & 63 & 48 \\ 
8 & 337920 &138240 & 70 & 52 \\ 
9 & 439296 &168960 & 76 & 55 \\ 
\bottomrule
\end{tabular} &
\begin{tabular}[t]{c cccc }
& \multicolumn{4}{c}{grid $24 \times 8^3$} \\
\toprule
$k$ & DOF\textsubscript{T} & DOF\textsubscript{F} & eDOFu-1D\textsubscript{T} & eDOFu-1D\textsubscript{F} \\
\midrule
1 &   405504 &  368640 &  50 &  42 \\
2 &   847872 &  688128 &  63 &  53 \\
3 &  1523712 & 1105920 &  75 &  63 \\
4 &  2482176 & 1622016 &  88 &  72 \\
5 &  3772416 & 2236416 & 101 &  80 \\
6 &  5443584 & 2949120 & 114 &  88 \\
7 &  7544832 & 3760128 & 127 &  96 \\
8 & 10125312 & 4669440 & 139 & 103 \\
\bottomrule
\end{tabular} \\
\end{tabular}
\caption{Taylor--Green vortex test case. 
Total number of cells (DOF\textsubscript{T}) and faces (DOF\textsubscript{F}) degrees of freedom.
For the sake of comparison with DNS data and other methods employing Cartesian element meshes we also provide
the equivalent number of degrees of freedom per direction for each velocity component.
We distinguish among cells (eDOFu-1D\textsubscript{T}) and faces (eDOFu-1D\textsubscript{F}) equivalent DOFs.
\label{tab:TGV_dofs}}
\end{table}

The user-defined threshold tolerance for tuning the local time step adaptation procedure
and the corresponding total number of time steps ($N_{\delta t}$) are reported in Table~\ref{tab:TGV_time_adap}.
The minimum and maximum time steps encountered along the time integration history are also tabulated,
and the minimum step size is more than an order of magnitude smaller than the maximum. 
Thanks to the combination of fully implicit time integration and local time step adaptation,
the number of time steps required to complete the simulation is relatively small.
It is possible to appreciate that increasing the polynomial degree
of the HHO discretization leads to an increased number of time steps, 
even if the threshold tolerance is kept fixed.
This behavior can be motivated considering that the timescales of smaller vortical structures are smaller
and that increasing the spatial accuracy allows to resolve higher-and-higher wavenumbers. 
We can indeed confirm that the time integration marches with the smaller time steps right after the peak of dissipation. 
The steep rise of the number of time steps suggests that, for the chosen value of $\tola$, 
the temporal error is still negligible compared to the spatial error and does not limit the overall accuracy.
\begin{table}[!t]
\centering
\small
\begin{tabular}{cc}
\begin{tabular}[t]{ccccc}
\multicolumn{5}{c}{grid $24\times 4^{3}$} \\
\toprule
$k$     & $\tola$ 	& $N_{\delta t}$ & $\delta t_{\rm MIN}$ & $\delta t_{\rm MAX}$ \\
\toprule
1       & 1e-04     & 24 			 & 4.91e-1              & 1.50e-0  \\
2       & 1e-04     & 34 			 & 4.14e-1              & 1.05e-0  \\
3       & 1e-04     & 45  			 & 2.84e-1              & 1.12e-0  \\
4       & 1e-04     & 56 			 & 1.67e-1              & 1.00e-0  \\
5       & 1e-04     & 69 			 & 1.35e-1              & 8.44e-1  \\
6       & 1e-04     & 78 			 & 1.14e-1              & 9.69e-1  \\
7       & 1e-04     & 86 			 & 9.67e-2              & 9.68e-1  \\
8       & 3.16e-05  & 120 			 & 6.35e-2              & 8.23e-1  \\
9       & 1e-05     & 172 			 & 4.22e-2              & 7.46e-1  \\
\bottomrule
\end{tabular}
&
\begin{tabular}[t]{ccccc}
\multicolumn{5}{c}{grid $24\times 8^{3}$} \\
\toprule
$k$     & $\tola$ 	& $N_{\delta t}$ & $\delta t_{\rm MIN}$ & $\delta t_{\rm MAX}$ \\
\toprule
1       & 1e-04     & 39 			 & 3.68e-1              & 7.77e-1  \\
2       & 1e-05     & 111			 & 9.51e-2              & 5.44e-1  \\
3       & 1e-05     & 133 			 & 6.90e-2              & 7.56e-1  \\
4       & 1e-05     & 155			 & 5.55e-2              & 7.46e-1  \\
5       & 1e-05     & 168			 & 4.72e-2              & 7.46e-1  \\
6       & 1e-05     & 180 			 & 3.99e-2              & 7.46e-1  \\
7       & 1e-05     & 193 			 & 4.32e-2              & 7.46e-1  \\
8       & 1e-05     & 204 			 & 4.19e-2              & 6.01e-1  \\
\bottomrule
\end{tabular} \\
\end{tabular}
\caption{Taylor--Green vortex test case, ESDIRK-HHO formulation with fifth order accuracy in time. 
User-defined adaptation trigger tolerance ($\tola$), total number of time steps ($N_{\delta t}$), minimum ($\delta t_{\rm MIN}$) 
and maximum step size ($\delta t_{\rm MAX}$) are reported for each combination of computational grid and polynomial degree $k$. \label{tab:TGV_time_adap}}
\end{table}

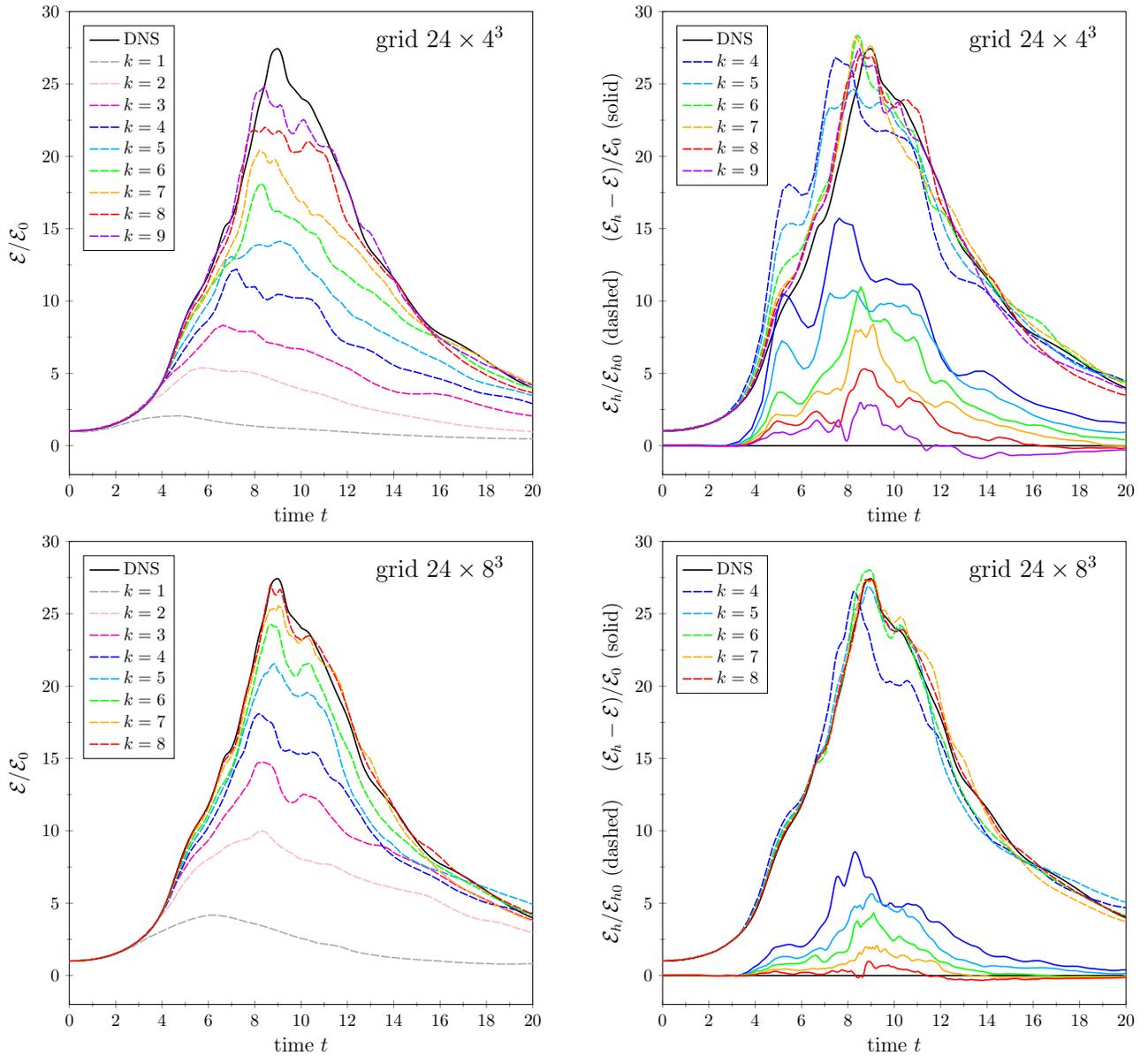
\begin{figure}[!t]
    \centering
\begin{tikzpicture}[scale=0.75]
\begin{axis}    [
                unbounded coords=jump,
                scale only axis=true,
                width=0.5\textwidth,
                height=0.5\textwidth,
                xlabel=time $t$,
                ylabel= $\mathcal{E} / \mathcal{E}_0$,
                label style={anchor=near ticklabel, font=\large},
                ticks=both,
                minor tick num=1,
                tick pos=left,
                tick align=center,
                xmin=0,
                xmax=20,
                ymin=-2,
                ymax=30,
                legend style={cells={anchor=west}},
                legend pos=north west,
                ]
\addplot    [
            thick,
            mark = none
            ]
            table[
            x expr = {\thisrowno{0}},
            y expr = {\thisrowno{3}/0.374989843653},
            header=false]
            {TGV_dns.dat};
\addplot    [
            thick,
            dash pattern=on 5pt off 1pt,
            mark = none,
            color = col_p1
            ]
            table[
            x expr = {\thisrowno{0}},
            y expr = {\thisrowno{3}},
            header=false]
            {TGV_g4_p1.dat};
\addplot    [
            thick,
            dash pattern=on 5pt off 1pt,
            mark = none,
            color = col_p2
            ]
            table[
            x expr = {\thisrowno{0}},
            y expr = {\thisrowno{3}},
            header=false]
            {TGV_g4_p2.dat};
\addplot    [
            thick,
            dash pattern=on 5pt off 1pt,
            mark = none,
            color = col_p3
            ]
            table[
            x expr = {\thisrowno{0}},
            y expr = {\thisrowno{3}},
            header=false]
            {TGV_g4_p3.dat};
\addplot    [
            thick,
            dash pattern=on 5pt off 1pt,
            mark = none,
            color = col_p4
            ]
            table[
            x expr = {\thisrowno{0}},
            y expr = {\thisrowno{3}},
            header=false]
            {TGV_g4_p4.dat};
\addplot    [
            thick,
            dash pattern=on 5pt off 1pt,
            mark = none,
            color = col_p5
            ]
            table[
            x expr = {\thisrowno{0}},
            y expr = {\thisrowno{3}},
            header=false]
            {TGV_g4_p5.dat};
\addplot    [
            thick,
            dash pattern=on 5pt off 1pt,
            mark = none,
            color = col_p6
            ]
            table[
            x expr = {\thisrowno{0}},
            y expr = {\thisrowno{3}},
            header=false]
            {TGV_g4_p6.dat};
\addplot    [
            thick,
            dash pattern=on 5pt off 1pt,
            mark = none,
            color = col_p7
            ]
            table[
            x expr = {\thisrowno{0}},
            y expr = {\thisrowno{3}},
            header=false]
            {TGV_g4_p7.dat};
\addplot    [
            thick,
            dash pattern=on 5pt off 1pt,
            mark = none,
            color = col_p8
            ]
            table[
            x expr = {\thisrowno{0}},
            y expr = {\thisrowno{3}},
            header=false]
            {TGV_g4_p8.dat};
\addplot    [
            thick,
            dash pattern=on 5pt off 1pt,
            mark = none,
            color = col_p9
            ]
            table[
            x expr = {\thisrowno{0}},
            y expr = {\thisrowno{3}},
            header=false]
            {TGV_g4_p9.dat};
\node[anchor=south, font=\Large] at (rel axis cs:0.8,0.9) {grid $24\times 4^{3}$};
\legend {DNS, $k=1$, $k=2$, $k=3$, $k=4$, $k=5$, $k=6$, $k=7$, $k=8$, $k=9$}
\end{axis}
\end{tikzpicture}
\hspace{5mm}
\begin{tikzpicture}[scale=0.75]
\begin{axis}    [
                unbounded coords=jump,
                scale only axis=true,
                width=0.5\textwidth,
                height=0.5\textwidth,
                xlabel=time $t$,
                ylabel= $\mathcal{E}_{h} / {\mathcal{E}_{h}}_0$ (dashed) \,\,\,\, $(\mathcal{E}_{h} -\mathcal{E} ) / {\mathcal{E}}_0$ (solid),
                label style={anchor=near ticklabel, font=\large},
                ticks=both,
                minor tick num=1,
                tick pos=left,
                tick align=center,
                xmin=0,
                xmax=20,
                ymin=-2,
                ymax=30,
                legend style={cells={anchor=west}},
                legend pos=north west,
                ]
\addplot    [
            thick,
            mark = none
            ]
            table[
            x expr = {\thisrowno{0}},
            y expr = {\thisrowno{3}/0.374989843653},
            header=false]
            {TGV_dns.dat};
\addplot    [
            thick,
            dash pattern=on 5pt off 1pt,
            mark = none,
            color = col_p4
            ]
            table[
            x expr = {\thisrowno{0}},
            y expr = {\thisrowno{2}},
            header=false]
            {TGV_g4_p4.dat};
\addplot    [
            thick,
            dash pattern=on 5pt off 1pt,
            mark = none,
            color = col_p5
            ]
            table[
            x expr = {\thisrowno{0}},
            y expr = {\thisrowno{2}},
            header=false]
            {TGV_g4_p5.dat};
\addplot    [
            thick,
            dash pattern=on 5pt off 1pt,
            mark = none,
            color = col_p6
            ]
            table[
            x expr = {\thisrowno{0}},
            y expr = {\thisrowno{2}},
            header=false]
            {TGV_g4_p6.dat};
\addplot    [
            thick,
            dash pattern=on 5pt off 1pt,
            mark = none,
            color = col_p7
            ]
            table[
            x expr = {\thisrowno{0}},
            y expr = {\thisrowno{2}},
            header=false]
            {TGV_g4_p7.dat};
\addplot    [
            thick,
            dash pattern=on 5pt off 1pt,
            mark = none,
            color = col_p8
            ]
            table[
            x expr = {\thisrowno{0}},
            y expr = {\thisrowno{2}},
            header=false]
            {TGV_g4_p8.dat};
\addplot    [
            thick,
            dash pattern=on 5pt off 1pt,
            mark = none,
            color = col_p9
            ]
            table[
            x expr = {\thisrowno{0}},
            y expr = {\thisrowno{2}},
            header=false]
            {TGV_g4_p9.dat};
\addplot    [
   			thick,
    		mark=none
			]
			coordinates {
    		(0,0) (20,0)
			};
\addplot    [
            thick,
            mark = none,
            color = col_p4
            ]
            table[
            x expr = {\thisrowno{0}},
            y expr = {\thisrowno{4}},
            header=false]
            {TGV_g4_p4.dat};
\addplot    [
            thick,
            mark = none,
            color = col_p5
            ]
            table[
            x expr = {\thisrowno{0}},
            y expr = {\thisrowno{4}},
            header=false]
            {TGV_g4_p5.dat};
\addplot    [
            thick,
            mark = none,
            color = col_p6
            ]
            table[
            x expr = {\thisrowno{0}},
            y expr = {\thisrowno{4}},
            header=false]
            {TGV_g4_p6.dat};
\addplot    [
            thick,
            mark = none,
            color = col_p7
            ]
            table[
            x expr = {\thisrowno{0}},
            y expr = {\thisrowno{4}},
            header=false]
            {TGV_g4_p7.dat};
\addplot    [
            thick,
            mark = none,
            color = col_p8
            ]
            table[
            x expr = {\thisrowno{0}},
            y expr = {\thisrowno{4}},
            header=false]
            {TGV_g4_p8.dat};
\addplot    [
            thick,
            mark = none,
            color = col_p9
            ]
            table[
            x expr = {\thisrowno{0}},
            y expr = {\thisrowno{4}},
            header=false]
            {TGV_g4_p9.dat};
\node[anchor=south, font=\Large] at (rel axis cs:0.8,0.9) {grid $24\times 4^{3}$};
\legend {DNS, $k=4$, $k=5$, $k=6$, $k=7$, $k=8$, $k=9$}
\end{axis}
\end{tikzpicture}
\begin{tikzpicture}[scale=0.75]
\begin{axis}    [
                unbounded coords=jump,
                scale only axis=true,
                width=0.5\textwidth,
                height=0.5\textwidth,
                xlabel=time $t$,
                ylabel= $\mathcal{E} / \mathcal{E}_0$,
                label style={anchor=near ticklabel, font=\large},
                ticks=both,
                minor tick num=1,
                tick pos=left,
                tick align=center,
                xmin=0,
                xmax=20,
                ymin=-2,
                ymax=30,
                legend style={cells={anchor=west}},
                legend pos=north west,
                ]
\addplot    [
            thick,
            mark = none
            ]
            table[
            x expr = {\thisrowno{0}},
            y expr = {\thisrowno{3}/0.374989843653},
            header=false]
            {TGV_dns.dat};
\addplot    [
            thick,
            dash pattern=on 5pt off 1pt,
            mark = none,
            color = col_p1
            ]
            table[
            x expr = {\thisrowno{0}},
            y expr = {\thisrowno{3}},
            header=false]
            {TGV_g8_p1.dat};
\addplot    [
            thick,
            dash pattern=on 5pt off 1pt,
            mark = none,
            color = col_p2
            ]
            table[
            x expr = {\thisrowno{0}},
            y expr = {\thisrowno{3}},
            header=false]
            {TGV_g8_p2.dat};
\addplot    [
            thick,
            dash pattern=on 5pt off 1pt,
            mark = none,
            color = col_p3
            ]
            table[
            x expr = {\thisrowno{0}},
            y expr = {\thisrowno{3}},
            header=false]
            {TGV_g8_p3.dat};
\addplot    [
            thick,
            dash pattern=on 5pt off 1pt,
            mark = none,
            color = col_p4
            ]
            table[
            x expr = {\thisrowno{0}},
            y expr = {\thisrowno{3}},
            header=false]
            {TGV_g8_p4.dat};
\addplot    [
            thick,
            dash pattern=on 5pt off 1pt,
            mark = none,
            color = col_p5
            ]
            table[
            x expr = {\thisrowno{0}},
            y expr = {\thisrowno{3}},
            header=false]
            {TGV_g8_p5.dat};
\addplot    [
            thick,
            dash pattern=on 5pt off 1pt,
            mark = none,
            color = col_p6
            ]
            table[
            x expr = {\thisrowno{0}},
            y expr = {\thisrowno{3}},
            header=false]
            {TGV_g8_p6.dat};
\addplot    [
            thick,
            dash pattern=on 5pt off 1pt,
            mark = none,
            color = col_p7
            ]
            table[
            x expr = {\thisrowno{0}},
            y expr = {\thisrowno{3}},
            header=false]
            {TGV_g8_p7.dat};
\addplot    [
            thick,
            dash pattern=on 5pt off 1pt,
            mark = none,
            color = col_p8
            ]
            table[
            x expr = {\thisrowno{0}},
            y expr = {\thisrowno{3}},
            header=false]
            {TGV_g8_p8.dat};
\node[anchor=south, font=\Large] at (rel axis cs:0.8,0.9) {grid $24\times 8^{3}$};
\legend {DNS, $k=1$, $k=2$, $k=3$, $k=4$, $k=5$, $k=6$, $k=7$, $k=8$, $k=9$}
\end{axis}
\end{tikzpicture}
\hspace{5mm}
\begin{tikzpicture}[scale=0.75]
\begin{axis}    [
                unbounded coords=jump,
                scale only axis=true,
                width=0.5\textwidth,
                height=0.5\textwidth,
                xlabel=time $t$,
                ylabel= $\mathcal{E}_{h} / {\mathcal{E}_{h}}_0$ (dashed) \,\,\,\, $(\mathcal{E}_{h} -\mathcal{E} ) / {\mathcal{E}}_0$ (solid),
                label style={anchor=near ticklabel, font=\large},
                ticks=both,
                minor tick num=1,
                tick pos=left,
                tick align=center,
                xmin=0,
                xmax=20,
                ymin=-2,
                ymax=30,
                legend style={cells={anchor=west}},
                legend pos=north west,
                ]
\addplot    [
            thick,
            mark = none
            ]
            table[
            x expr = {\thisrowno{0}},
            y expr = {\thisrowno{3}/0.374989843653},
            header=false]
            {TGV_dns.dat};
\addplot    [
            thick,
            dash pattern=on 5pt off 1pt,
            mark = none,
            color = col_p4
            ]
            table[
            x expr = {\thisrowno{0}},
            y expr = {\thisrowno{2}},
            header=false]
            {TGV_g8_p4.dat};
\addplot    [
            thick,
            dash pattern=on 5pt off 1pt,
            mark = none,
            color = col_p5
            ]
            table[
            x expr = {\thisrowno{0}},
            y expr = {\thisrowno{2}},
            header=false]
            {TGV_g8_p5.dat};
\addplot    [
            thick,
            dash pattern=on 5pt off 1pt,
            mark = none,
            color = col_p6
            ]
            table[
            x expr = {\thisrowno{0}},
            y expr = {\thisrowno{2}},
            header=false]
            {TGV_g8_p6.dat};
\addplot    [
            thick,
            dash pattern=on 5pt off 1pt,
            mark = none,
            color = col_p7
            ]
            table[
            x expr = {\thisrowno{0}},
            y expr = {\thisrowno{2}},
            header=false]
            {TGV_g8_p7.dat};
\addplot    [
            thick,
            dash pattern=on 5pt off 1pt,
            mark = none,
            color = col_p8
            ]
            table[
            x expr = {\thisrowno{0}},
            y expr = {\thisrowno{2}},
            header=false]
            {TGV_g8_p8.dat};
\addplot    [
   			thick,
    		mark=none
			]
			coordinates {
    		(0,0) (20,0)
			};
\addplot    [
            thick,
            mark = none,
            color = col_p4
            ]
            table[
            x expr = {\thisrowno{0}},
            y expr = {\thisrowno{4}},
            header=false]
            {TGV_g8_p4.dat};
\addplot    [
            thick,
            mark = none,
            color = col_p5
            ]
            table[
            x expr = {\thisrowno{0}},
            y expr = {\thisrowno{4}},
            header=false]
            {TGV_g8_p5.dat};
\addplot    [
            thick,
            mark = none,
            color = col_p6
            ]
            table[
            x expr = {\thisrowno{0}},
            y expr = {\thisrowno{4}},
            header=false]
            {TGV_g8_p6.dat};
\addplot    [
            thick,
            mark = none,
            color = col_p7
            ]
            table[
            x expr = {\thisrowno{0}},
            y expr = {\thisrowno{4}},
            header=false]
            {TGV_g8_p7.dat};
\addplot    [
            thick,
            mark = none,
            color = col_p8
            ]
            table[
            x expr = {\thisrowno{0}},
            y expr = {\thisrowno{4}},
            header=false]
            {TGV_g8_p8.dat};
\node[anchor=south, font=\Large] at (rel axis cs:0.8,0.9) {grid $24\times 8^{3}$};
\legend {DNS, $k=4$, $k=5$, $k=6$, $k=7$, $k=8$}
\end{axis}
\end{tikzpicture}
    \caption{Taylor--Green vortex test case, ESDIRK-HHO formulation with fifth order accuracy in time and $k=1,\dots,9$. Adaptive local time stepping is employed.
             \emph{Left}: relative enstrophy $\frac{\mathcal{E}}{\mathcal{E}_0}$ (dashed line). \emph{Right}: Relative total dissipation $\frac{\mathcal{E}_h}{{\mathcal{E}_h}_0}$ (dotted line) and relative numerical dissipation $\frac{\mathcal{E}_h - \mathcal{E}}{\mathcal{E}_0}$ (solid line), see text for details. 
The notation $\mathcal{E}_0=\mathcal{E}|_{t_0}$ and ${\mathcal{E}_{h}}_0 =\mathcal{E}_h|_{t_0}$ is employed. \label{TGVenstro}}
\end{figure}

In Figure \ref{TGVenstro}, besides comparing the relative enstrophy $\frac{\mathcal{E}}{\mathcal{E}_0}$ with DNS data,
we show the relative total dissipation $\frac{\mathcal{E}_h}{{\mathcal{E}_h}_0}$ and the relative numerical dissipation
$\frac{\mathcal{E}_h - \mathcal{E}}{\mathcal{E}_0}$ for a subset of the runs.
We simply consider the most relevant results, that is the most accurate ones, in order to show the influence of
refining the grid and increasing the polynomial degree.
Notice that, for sake of conciseness, the notation 
$\mathcal{E}_{0}=\mathcal{E}|_{t_0}$ and $\mathcal{E}_{h0}=\mathcal{E}_{h}|_{t_0}$ is employed.
We remark that, when considering relative quantities (with respect to initial time $t=0$), we can alternatively refer to enstrophy or dissipation,
and, in case of DNS, total and physical dissipation are basically the same.
As reported in the literature, the enstrophy behavior is the most challenging to replicate.
The relative total dissipation shows good agreement with the DNS relative enstrophy, even at relatively low polynomial degrees, confirming the
robustness of the ESDIRK-HHO formulation in under resolved computations.
However, we remark that, on the coarser grid, the total dissipation curve is
shifted to the left compared to DNS data, even at the highest $k=9$ polynomial degree.
Note that, in the enstrophy increasing phase, $k=7,8,9$ computations are in good agreement in terms of total dissipation,
suggesting that much higher polynomial degrees would be required to properly replicate DNS.
The behaviour of the numerical dissipation outlines the benefit of increasing the polynomial degree and refining the grid.
We remark that, on the finer grid at $k=8$, the maximum relative numerical dissipation is smaller than one,
which, compared with the relative enstrophy value at peak, implies that the evolution of
kinetic energy shows a 3\% discrepancy with respect to a fully resolved computation.

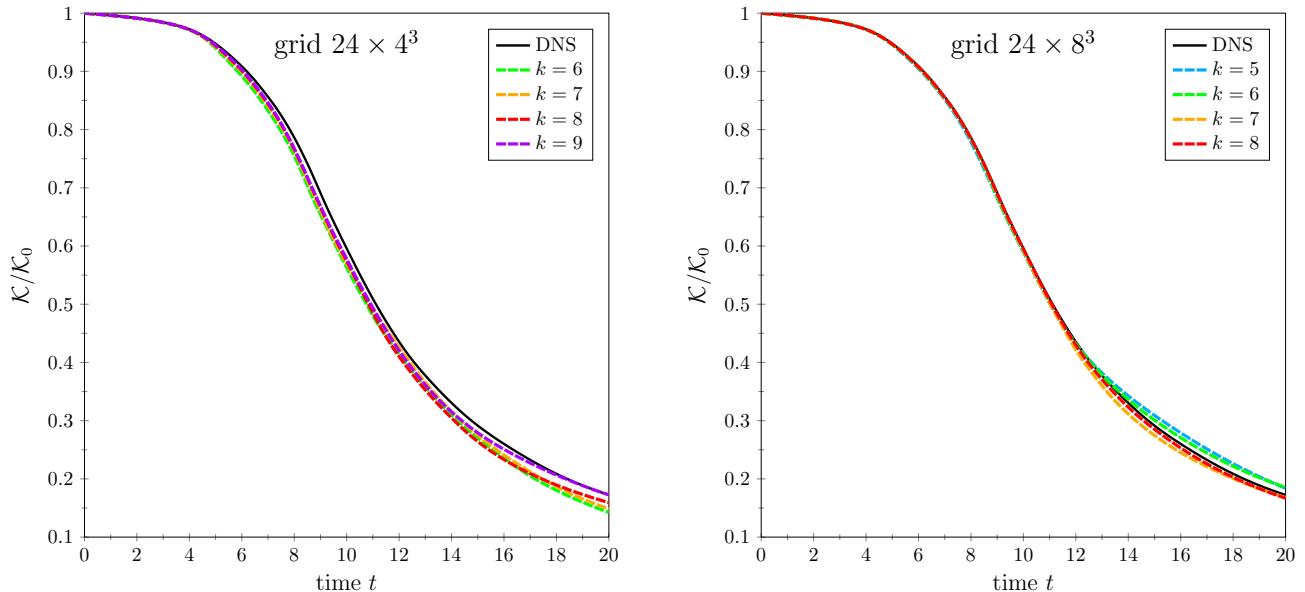
\begin{figure}[!t]
    \centering
\begin{tikzpicture}[scale=0.75]
\begin{axis}    [
                unbounded coords=jump,
                scale only axis=true,
                width=0.5\textwidth,
                height=0.5\textwidth,
                xlabel=time $t$,
                ylabel= $\mathcal{K} / \mathcal{K}_0$,
                label style={anchor=near ticklabel, font=\large},
                ticks=both,
                minor tick num=1,
                tick pos=left,
                tick align=center,
                xmin=0,
                xmax=20,
                ymin=0.1,
                ymax=1,
                legend style={cells={anchor=west}},
                legend pos=north east,
                ]
\addplot    [
            very thick,
            mark = none
            ]
            table[
            x expr = {\thisrowno{0}},
            y expr = {\thisrowno{1}/0.125},
            header=false]
            {TGV_dns.dat};
\addplot    [
            ultra thick,
            dash pattern=on 5pt off 1pt,
            mark = none,
            color = col_p6
            ]
            table[
            x expr = {\thisrowno{0}},
            y expr = {\thisrowno{1}},
            header=false]
            {TGV_g4_p6.dat};
\addplot    [
            ultra thick,
            dash pattern=on 5pt off 1pt,
            mark = none,
            color = col_p7
            ]
            table[
            x expr = {\thisrowno{0}},
            y expr = {\thisrowno{1}},
            header=false]
            {TGV_g4_p7.dat};
\addplot    [
            ultra thick,
            dash pattern=on 5pt off 1pt,
            mark = none,
            color = col_p8
            ]
            table[
            x expr = {\thisrowno{0}},
            y expr = {\thisrowno{1}},
            header=false]
            {TGV_g4_p8.dat};
\addplot    [
            ultra thick,
            dash pattern=on 5pt off 1pt,
            mark = none,
            color = col_p9
            ]
            table[
            x expr = {\thisrowno{0}},
            y expr = {\thisrowno{1}},
            header=false]
            {TGV_g4_p9.dat};
\node[anchor=south, font=\Large] at (rel axis cs:0.5,0.9) {grid $24\times 4^{3}$};
\legend {DNS, $k=6$, $k=7$, $k=8$, $k=9$}
\end{axis}
\end{tikzpicture}
\hspace{5mm}
\begin{tikzpicture}[scale=0.75]
\begin{axis}    [
                unbounded coords=jump,
                scale only axis=true,
                width=0.5\textwidth,
                height=0.5\textwidth,
                xlabel=time $t$,
                ylabel= $\mathcal{K} / \mathcal{K}_0$,
                label style={anchor=near ticklabel, font=\large},
                ticks=both,
                minor tick num=1,
                tick pos=left,
                tick align=center,
                xmin=0,
                xmax=20,
                ymin=0.1,
                ymax=1,
                legend style={cells={anchor=west}},
                legend pos=north east,
                ]
\addplot    [
            very thick,
            mark = none
            ]
            table[
            x expr = {\thisrowno{0}},
            y expr = {\thisrowno{1}/0.125},
            header=false]
            {TGV_dns.dat};
\addplot    [
            ultra thick,
            dash pattern=on 5pt off 1pt,
            mark = none,
            color = col_p5
            ]
            table[
            x expr = {\thisrowno{0}},
            y expr = {\thisrowno{1}},
            header=false]
            {TGV_g8_p5.dat};
\addplot    [
            ultra thick,
            dash pattern=on 5pt off 1pt,
            mark = none,
            color = col_p6
            ]
            table[
            x expr = {\thisrowno{0}},
            y expr = {\thisrowno{1}},
            header=false]
            {TGV_g8_p6.dat};
\addplot    [
            ultra thick,
            dash pattern=on 5pt off 1pt,
            mark = none,
            color = col_p7
            ]
            table[
            x expr = {\thisrowno{0}},
            y expr = {\thisrowno{1}},
            header=false]
            {TGV_g8_p7.dat};
\addplot    [
            ultra thick,
            dash pattern=on 5pt off 1pt,
            mark = none,
            color = col_p8
            ]
            table[
            x expr = {\thisrowno{0}},
            y expr = {\thisrowno{1}},
            header=false]
            {TGV_g8_p8.dat};
\node[anchor=south, font=\Large] at (rel axis cs:0.5,0.9) {grid $24\times 8^{3}$};
\legend {DNS, $k=5$, $k=6$, $k=7$, $k=8$}
\end{axis}
\end{tikzpicture}
    \caption{Taylor--Green vortex test case, ESDIRK-HHO formulation with fifth order accuracy in time. Adaptive local time stepping is employed.
             \emph{Left and right}: relative kinetic energy evolution $\frac{\mathcal{K}}{\mathcal{K}_{0}}$ on the coarse $24*4^3$ and the fine $24*8^3$ grids. 
                                  The notation $\mathcal{K}_0=\mathcal{K}|_{t_0}$ is employed. \label{TGVenergy}}
\end{figure}

In Figure \ref{TGVenergy} we compare the relative kinetic energy $\mathcal{K}/\mathcal{K}_{0}$
time evolution with DNS data for the most accurate results on each mesh. Notice that $\mathcal{K}_{0}=\mathcal{K}|_{t_{0}}$.
On the coarser grid, the left shift of the dissipation observed in the entrophy analysis induces an earlier than expected decrease of the kinetic energy.
On the finer grid, the evolution of kinetic energy replicates DNS data in the entrophy increasing phase 
while some discrepancies can still be appreciated in the entrophy decreasing phase.
Nevertheless, the improvements gained increasing the polynomial degree 
suggests that a full resolution of energy cascade phenomena is not out of reach.

Overall, considering that the computational meshes are rather coarse and unstructured, which is probably suboptimal
for this test case, the results are satisfactory. 
Note that, despite the number of degrees of freedom is a fraction of the amount employed to achieve DNS data, 
the trend towards DNS is confirmed by increasingly precise replication of the enstrophy behavior.  
We remark that the focus on tetrahedral elements meshes is motivated by the observation that four nodes tetrahedral
cells have planar faces, independently of the disposition of the nodes. 
We also remind that planar faces are essential to retain efficiency of static condensation 
and pressure robustness all together \cite{Botti.Di-Pietro:18,YEMMHHOC24}.
Accordingly, when tackling complex computational domains,
relying on simplicial meshes is a viable and straightforward choice in the context of hybrid formulations.
We also refer to Section~\ref{sec:remarkNonSimp} for a discussion on the application of ESDIRK-HHO to non-simplicial meshes.


\section{Conclusion}\label{sec:conclusion}

We provided extensive numerical validation, in both two and three space dimensions,
of an ESDIRK-HHO formulation of the incompressible Navier--Stokes equations that is capable of
reaching high-orders of accuracy in both space and time.
Relevant features are pressure-robustness, that is decoupling of the velocity error with respect to the pressure error,
and exact conservation of mass, with point-wise divergence free velocity fields.
The proposed implementation relies on $p$-multilevel solution strategies and static condensation
to alleviate the computational burden associated with fully implicit time marching strategies.
This combination has proved effective thanks to use of fully hybrid polynomial spaces, for both the velocity and the pressure.
Moreover, the time step can be locally adapted to improve accuracy of the time integration.

Robustness has been demonstrated tackling challenging test case and considering both
viscous dominated and convection-dominated flow regimes, up to the inviscid limit.
To corroborate the turbulence modelling capabilities claim, we performed the Taylor--Green Vortex problem
showing the ability to reach DNS like precision in the relevant flow features measures
employing coarse meshes and high-polynomial degrees.

\section*{Acknowledgements}

\noindent Daniele Di Pietro acknowledges funding by the European Union (ERC Synergy, NEMESIS, project number 101115663).
Views and opinions expressed are however those of the authors only and do not necessarily reflect those of the European Union or the European Research Council Executive Agency. Neither the European Union nor the granting authority can be held responsible for them.

\bibliographystyle{elsarticle-num}

\bibliography{hdivHHOtm.bib}







\end{document}